\input amstex
\magnification=\magstep1 \baselineskip=24pt
\font\eightit=cmti8
\font\eightbf=cmbx8

\vskip .3in

 \centerline{\bf      The Axiomatisation of Physics}
\centerline {by Joseph F.\ Johnson}
\baselineskip=12pt 
\centerline {Math Dept., Villanova Univ.\ }
\baselineskip=24pt
\vskip .2in

\centerline{\bf Abstract} \baselineskip=12pt 
\font\eightrm=cmr8
\eightrm

Analysing Quantum Measurement requires analysing 
the physics of amplification since amplification of phenomena from one scale to
another scale is essential to measurement.
There still remains the task of working this into an axiomatic logical 
structure, what should be the foundational status of the concepts of measurement 
and probability.  We argue that the concept of physical probability is a 
multi-scale phenomenon and as such, can be explicitly defined in terms of 
more fundamental physical concepts.  Thus Quantum Mechanics can be given a 
logically unexceptionable axiomatisation.  
We introduce a new definition of macroscopic observable which 
implements Bohr's insight that the observables of a measurement apparatus are 
classical in nature.  In particular, we obtain the usual non-abelian observables
 as limits of abelian, classical, observables.  
This is the essential step in Hilbert's Sixth Problem.

\baselineskip=24pt 
\rm

\centerline{\bf Introduction}

Hilbert's Sixth Problem is the axiomatisation of Physics.  In 1900 at the 
International Congress of Mathematicians in Paris, Hilbert proposed 23 problems 
the solutions of which, he predicted, would require major advances in 
mathematical technique and lead to major advances in mathematical knowledge.
Some few of these problems have been fulfilled neither of these expectations,
but most of them have amply repaid expectations.
The first six problems were about foundations, of logic, set theory, geometry, 
and Physics.

The Sixth Problem has not been considered one of his best.  
It has generally been considered sort of borderline.  This paper will 
argue that that is because it has been misunderstood and Hilbert's original 
focus, in its historical context, has been lost sight of.  The revolutionary 
developments in Physics that followed 1900 have only reinforced the prescience 
of Hilbert's own formulation of the difficulty in the foundations of Physics.
In fact the essence of the solution has been published, in various pieces, in 
the physics literature already, and only needs to be clarified and assembled.

From the standpoint of axiomatics, Physics has been in a crisis, or at least 
a mess, from 1927\plainfootnote*
{\eightrm \baselineskip=8pt 
As J.S. Bell put it in his last published article[3], `Surely, after 62 years, we should have an exact formulation of some serious 
part of quantum mechanics?  By ``exact'' I do not of course mean ``exactly 
true.''  I mean only that the theory should be fully formulated in 
mathematical terms, with nothing left to the discretion of the theoretical 
physicist \dots until workable approximations are needed in applications.
By ``serious'' I mean that some substantial fragment of physics should 
be covered.  Nonrelativistic ``particle'' quantum mechanics, perhaps 
with the inclusion of the electromagnetic field and a cut-off interaction, 
is serious enough.  For it covers `a large part of physics and the whole 
of chemistry.' I mean, too, by ``serious'' that ``apparatus'' should not 
be separated off from the rest of the world into black boxes, as if 
it were not made of atoms and ruled by quantum mechanics.'}
on: the problem is called that of Quantum Measurement.

So much so, that ever since then, it 
has often been questioned whether the axiomatic method is even appropriate 
for Physics.  (That, of course, is why Hilbert suggested that mathematicians 
should do the work on this problem.)  One of Hilbert's other foundational problems 
in mathematics has indeed turned out contrary to his conjecture: Goedel's 
famous undecidability theorem showed that Hilbert was too optimistic in the 
statement of his conjecture.  Now if in fact it is the case that the axiomatic 
method of Euclid, Archimedes, Newton, Hertz, Klein, and Hilbert is inappropriate
to Physics, that would be a claim whose definitive establishment could only be
accomplished by an axiomatic analysis of the foundations of Physics similar to
Goedel's.  This must be regarded as an open question, but it will be answered 
by this paper that no such fear is really justified:  
Quantum Mechanics does not introduce anything new 
in this regard in comparison to Classical Mechanics, there is nothing in 
Quantum Mechanics to stand in the way of an axiomatic foundation to Physics 
such as Newton and Hertz essayed.  Even more precisely, there is no need to 
change normal 18th century logic or philosophy.  It is possible to axiomatise 
Quantum Mechanics, at least, in the same way Newton and Hertz operated with 
Classical Mechanics.

We will neglect gravity, and assume that Schroedinger's equation is universally 
and precisely true, and is exactly linear and unitary.  One might wonder 
whether future developments in Physics will render these considerations irrelevant.
Many physicists have, privately, felt that the present axiomatic mess of 
Quantum Mechanics would just be rendered irrelevant, obsolete, by future 
developments in Physics.  The methods of this paper will retain some interest 
in any case: they 
are robust, and so would probably carry over 
to future modifications, even non-linear ones.  
Some physicists 
such as Weinberg, feel that the logical structure of Quantum Mechanics is so 
rigid, so incapable of slight modifications,\plainfootnote
*{\baselineskip=8pt \eightrm\eightit Dreams of a Final Theory, \eightrm New York, 1992, 
pp.\ 87-89, 211.}
 that it will persist in any 
future theory, even at the Planck scale or involving strings or branes.
Be that as it may, this paper will cover `a serious part of Physics.'

The solution to be offered will be essentially 
statistical mechanical in nature.  J.S. Bell$^3$ has influentially criticised many 
earlier proferred solutions of a statistical mechanical nature, we will carefully 
meet the important points he has raised.  Our claim is that the underlying physics 
of many earlier solutions, such as that of H. S. Green$^9$, Daneri--Loinger--Prosperi
$^{10}$,
Colemann--Hepp$^{11}$, is basically sound, and Bell's criticisms can be met by being 
axiomatically and logically careful about the development of these same 
physical concepts.  The underlying physics is that of amplification due to the 
coupling between a microscopic system with a larger assembly in a state of 
negative temperature.  In order to meet Bell's criticisms, we will define 
the concepts of probability and measurement in terms of fundamental undefined 
notions such as energy, space, and time.  Such explicit definitions are the only 
ingredients really missing from published solutions.  Bell's criticisms were 
substantially logical, foundational, and axiomatic in nature, so it should not 
be surprising that meeting them requires axiomatic care, even to indulging 
the axiomatic temperament. 

Other attempts (such as Lucien Hardy's) to axiomatise Quantum Mechanics or Quantum Information Theory 
are quite antagonistic to Hilbert's point of view. 

\centerline{\bf History}

In 1900 when Hilbert proposed this problem, the landscape of Physics was very 
different from now.  For a hundred years or more, physical science had been 
riven by completely incompatible theores: mechanics was incompatible with 
electricity, thermodynamics with both, and atomism (the contribution of chemists) 
and the theory of chemical structure seemed like an alien kind of magic.  Vitalism 
and phlogiston theories were tolerated. 
Each rival theory was put on its own axiomatic basis, and the difficulty was 
that of reconciling the competing theories and domains.
But steady progress was made towards 
overcoming these contradictions, due not to advances in mathematical technique 
so much as physical insights.  Lorentz and Poincare were just formulating 
the precise way in which the symmetries of Maxwell's equations were 
incompatible with the symmetries of Newton's equations, and although this still 
left electrodynamics in a state of crisis, it was progress.  Maxwell and 
Boltzmann had also made serious progress on explaining heat and thermodyanmics 
in terms of statistical properties of the Newtonian mechanics of large 
assemblages of billiard balls of the same size as atoms.  Although this still 
left Statistical Mechanics in a state of crisis because of the problems with 
irreversibility and ergodicity, it encouraged Hilbert to pose the problem.  

Perhaps the reason Hilbert was still optimistic was because within mathematics 
itself, similar difficulties of competing domains had been triumphantly overcome.
In 1600, only geometry was axiomatic, and number theory and algebra were so 
amorphous that Fermat suffered from a writer's block about number theory which 
did not affect him when he wrote his geometric book.  By 1700, the new analysis 
had evolved with its own new undefined concepts, and without any axiomatic order
or organisation, and so the three different fields of Geometry, Algebra, and 
Analysis had the same sort of incompatibility and domain problems that afflicted 
Physics later.  But the heroic labours of Cauchy, Weierstrass, Grassmann, and 
Cantor put an end to this gradually.  It was called the arithmetisation of 
Analysis: the vague concepts of analysis were given precise definitions in 
terms of the concepts of arithmetic, algebra, and so on.  Klein contributed 
by unifying non-Euclidean geometry with Euclidean geometry, thus removing 
the mysticism and philosophical opposition to non-Euclidean geometry.  
Hilbert and others continued by a formal axiomatisation of Euclidean geometry 
(as opposed to Euclid's own, which was informal and seemed to require diagrams).
Peano, Cantor, and Frege gave clear and unambiguous definitions of `function', 
`variable', `number.'  Russell showed how all of Analysis needed for mathematical 
physics could be deduced from a dozen logical axioms and a few logical concepts.

Probability theory still stood aloof from this unification and had no axiomatic 
basis.  (Wittgenstein made a stab at it at the end of the Tractatus.)

By 1925 the landscape of Physics had completely changed.  But let us not get 
ahead of ourselves too much.  Let us try to see Hilbert's point of view in its 
context.  Although aware of Poincare's work, he did not focus on the crisis 
in electrodynamics.  He focussed on the logical foundations of Boltzmann's 
work.  

``6. Mathematical treatment of the axioms of physics
\baselineskip=8pt

\eightrm``The investigations on the foundations of geometry suggest the problem: To treat in the same manner, by means of axioms, those physical sciences in which mathematics plays an important part; in the first rank are the theory of probabilities and mechanics. 

\baselineskip=24pt \rm
This is amazingly prescient for 1900.  Special and General Relativity (to which Hilbert 
contributed Einstein's equation, a few days in advance of Einstein), soon succeeded 
in accomplishing the unification of Classical Mechanics with Electromagnetism, 
without producing any axiomatic messes.  But the controversy over the foundations of 
Statistical Mechanics is especially an axiomatic problem.  The task Maxwell and, later, 
Boltzmann, set themselves was to deduce thermodynamics from Hamiltonian Mechanics.
But in order to do this they introduced new methods which were powerful, but lacked 
proof, and when used without sufficient mathematical care, led to the following 
obvious contradiction: `entropy always increases' is in obvious contradiction to 
the time-reversal symmetry of Newton's laws of motions and to Poincare recurrence.
Worse, they introduced new physical concepts without adequate definition: the concepts 
of probability and mixed state.  (Now if one reformulates this, as has been done, 
in terms of information, this merely transfers the problem to a synonym: one has 
introduced the new concept of information without adequate physical definition.)

Sometimes an axiomatic analysis of foundations shows that one needs more primitive 
concepts and more explicit axioms about them than was realised.  For example,
in the axiomatic analysis of the concepts of geometry, where one found out that 
Euclid needed more concepts than he had specified: in order to eliminate Euclid's 
reliance on diagrams and intuition, Pasch, Hilbert, and others, had to introduce 
explicit foundational (hence undefined and primitive) concepts such as `between'.
Sometimes one finds out that concepts which everyone assumed had to be primitive
are in fact reducible to other concepts.  For example,
in analysis, where Cauchy and Weierstrass showed that fewer concepts were 
really needed, since undefined concepts like `limit' could after all be given 
definitions in terms of basic arithmetic concepts.  And Frege, after all, number 
did not need to be assumed, it could be defined, too.  So, Russell: it was no 
longer true that geometry was about a proprietary concept of point that could not 
be defined, or `order', with arithemtic about its own patented undefined notion 
of number, rather all these concepts previously thought to be foundational 
and undefined, could be defined in terms of `true', `proposition', etc.
So, phlogiston got eliminated by having temperature, heat, defined in terms of 
motion.  But, alas, probability and mixed state got introduced as new undefined 
concepts.  So an axiomatic analysis would investigate whether probability 
was needed, as a new primitive, or whether it could be defined in terms of the 
other undefined, primitive concepts.  Can the notion of mixed state be eliminated?
These questions are typical of the axiomatic temperament.

Hertz successfully displayed the axiomatic temperament when he reduced all laws 
of Newtonian mechanics to one, his own variational principle, and all primitive, 
undefined concepts to three.  After 1900, 
Minkowski, Hilbert, and Klein dabbled in Relativity to do the same.

Now let us fast forward to Schroedinger's development of Wave Mechanics 
(the relativistic equation he did not publish till years later, now usually called 
the Klein--Gordon--Schroedinger--Fock equation) in 1925.  All of Physics 
(except cosmology and gravity) is now explained by one equation.  Immediately, 
mathematicians and physicists under Hilbert's influence began the task of 
axiomatising the new mechanics.  Weyl, participating in a seminar with Debye 
and Schroedinger, was the first to publish an axiomatisation.  Hilbert, 
von Neumann, and Nordheim attempted one which is now forgotten.  Wigner and 
von Neumann continued these efforts and published one now regarded as definitive
of its type.  Dirac was by no means under Hilbert's influence, but under the 
influence of Sir James Jeans, Charles Galton Darwin, and Ralph Fowler.  He 
published his own, which is not expressed in the Klein--Hilbert style, but is 
often regarded as equivalent to the von Neumann--Wigner one.  

But the resulting system is not a system of axioms in Hilbert's sense (or in 
Euclid, Newton, Hertz, Klein, Russell, Bell, etc.)  Not because of its use 
of new concepts such as probability or measurement or observer or uncertainty.
Wigner analysed exactly what the new system's problem, from this perspective 
is, and gave it the name Quantum Duality.  This will be explained in detail 
in the next section of the paper, by way of anticipation, we will just say 
that the same physical situation can be analysed in two different ways by 
choosing to use some axioms and ignore others, or vice versa, and with 
different results.  The whole point of formalising a theory into axiomatic 
form is to eliminate the need for discretion, experience, good taste.\plainfootnote*
{\baselineskip=8pt \eightrm `There is nothing in the 
mathematics to tell what is ``system''  and what is ``apparatus'', nothing to tell
which natural processes have the special status of ``measurements''.  Discretion 
and good taste, born of experience, allow us to use quantum theory 
with marvelous success, despite the ambiguity of the concepts named above 
in quotation marks.  But it seems clear that in a serious fundamental 
formulation such concepts must be excluded.'  Bell, 
Beables for Quantum Field Theory. 1984 Aug 2, CERN-TH. 4035/84}\baselineskip=24pt

Can some basic 
concepts be eliminated by being defined in terms of the others? Is there more 
than one way to do this? What are the exact logical relationships between 
the different concepts and axioms?  Can some be deduced from others?
Can the dual overlap Wigner identified be eliminated?
The problem Hilbert first called attention to in 1900 becomes still more 
central in 1927 than anyone else would have thought.

Now let us rewind to Boltzmann, and his irreversibility, versus Zermelo and 
Poincare.  Which Hilbert was concerned with.  In the 20's, this problem was 
solved by Darwin and Fowler.  They put the existing statistical mechanics on 
a completely satisfactory, logically, foundation and even succeeded in solving 
some new problems with their methods.  Indeed, even by 1905, Jeans had outlined 
the common-sense but logically insightful resolution of this particular 
problem.\baselineskip=8pt \plainfootnote
*{\baselineskip=8pt \eightrm Some writers have interpreted this to mean that \eightit H \eightrm will continually decrease, until it reaches a minimum value, and will then retain that value for ever after.  A motion of this kind would, however, be dynamically irreversible, and therefore inconsistent with the dynamical equations of motion from which it professes to have been deduced.  As will appear later, the truth is that we have at this point reached the limit within which the assumption of molecular chaos lead to accurate results.  The motion is, in point of fact, strictly reversible, and the apparent irreversibility is merely an illusion introduced by the imperfections of the statistical method.  Jeans, \eightit The Dynamical Theory of Gases\eightrm, second edition, London, 1916, p.\ 38.} \baselineskip=24pt
 Again by way of anticipation, we will just outline the solution, 
which eliminates the need for the concept of probability.  Statistical 
Mechanics is the study of the statistical properties of the `normal' state
(or trajectory), and Jeans gave a reasonable definition of what he meant by 
`normal'.  Boltzmann's deduction of irreversibility takes place within a 
statistical model, which makes certain approximations compared to the exact 
Hamiltonian model.  That deductions from an approximation will be of limited validity 
does not deserve to be considered a crisis or paradox!

Next, let us pass to Hilbert's concern with the use of probabilities in 
Boltzmann's deductions.  Hilbert wisely pointed to two different aspects of 
probability.  The first is the notion of physical probability, as used by 
Boltzmann in his very formulations.  Parallel to this would be to clarify 
the mathematical definition of probability as used in the mathematical 
techniques used by Boltzmann and Maxwell in their deductions.  So Hilbert 
called for ultimately defining, if possible, and axiomatising a new undefined 
primitive, if necessary, of physical probability in such a way that it could 
be used as Boltzmann used it.  And as a useful preliminary to this, 
clarifying \it within mathematics merely\rm, the axiomatic status or 
definition of the mathematical concept of probability.

Borel and Levy resisted this to the end of their days, just as did Poincare 
resist Hilbert's logical aims.  Wiener, however, in 1922, following Frechet, 
used measure-theoretic notions of Borel and Lebesgue to rigourously 
formulate the notions of probabilites used heuristically by Einstein in 
the theory of Brownian Movement.  Then Kolmogoroff systematised this and 
published the influential axiomatisation of probability now well known.
From a Hilbertian point of view, what Kolmogoroff did was exactly what 
he asked for as a preliminary to the ultimate goal.  
But it should not be called, from a Hilbertian point of view, an 
axiomatisation of probability: it is quite the other possibility that 
Kolmogoroff succeeded in establishing: an explicit definition of mathematical 
probability in terms of prior, already defined, foundational concepts: those 
of measure.  Hence the explicit definition of (mathematical) probability 
is accomplished, and Kolmogoroff showed that the entire branch of mathematics 
known as probability theory was incorporated in the grandly unified structure, 
going back to those dozen axioms of logic\dots and no new undefined, primitive 
concept or axiom was required.  Within the scope and aims of the Hilbert problem, 
Kolmogoroff showed that the law of large numbers, say, was no different from 
the theorem 1+1=2 in \it Principia Mathematica\rm.  This eliminated the mysticism
and philosophy from the mathematical theory of probability, and was an essential 
step in avoiding seeming paradoxes involving continuous probabilities in 
geometric settings.  The theory of stochastic processes could not have 
developed in the continuous time setting without this.  (But Kolmogoroff 
well knew that it did not solve the ultimate goal of the axiomatic analysis of 
the notion of physical probability and did not solve the problem of giving 
a logically unexceptionable definition of the concept of physical probability.
This difficulty will be exposited in detail in a later part of this paper.)

Even after this accomplishment, 
it could be wondered whether the physical concept of probability was amenable 
to something similar, or would require new, Bohrian complementarity style 
axioms or new quantum logic axioms or goodness knows what else, or even 
the abandonment of the axiomatic method in Physics.  The point of this paper 
is to show that the concept of probability in Physics can be defined 
and the measurement axioms involving it be derived from more fundamental 
axioms and concepts, not involving the notions of measurement or probability
or information or observation.

We should not postulate a measure.  We must derive the measure from physical properties
of the system.  We should not introduce subjective notions (such as information) in 
order to derive the measure.  We may use the notions introduced by Poincare into 
`general dynamics' (such as stability or density or `almost everywhere').  These notions 
are mere logical combinations of physical notions and do not introduce anything new.

Perhaps it is appropriate to remark here that philosophical commitments should 
not be allowed to trump physics and logic.  If more than one solution is possible 
which satisfies the requirements of agreement with experimental results and 
the requirements of logic, then one could let philosophical predilections, like 
a bias in favour of positivism, help one choose between possible 
solutions.  But not otherwise.  Twentieth century physicists are more unanimous 
in philosophical commitments than they used to be, and this is not necessarily 
a good thing.  There are a variety of respectable philosophical positions 
even \it vis-a-vis\rm\  science, which have existed or are possible, and if there 
is a solution which satisfies the requirements of experiment, logic, and is 
consistent with any such respectable philosophy, then rival philosophical 
commitments should not be allowed much weight.  Currently, monism is widely 
accepted by scientists, but dualism used to be more current, for example, 
with Newton, Leibnitz, and Hertz\plainfootnote*
{\baselineskip=8pt \eightrm In the text we take the natural precaution of expressly limiting the range of our mechanics to inanimate nature; how far its laws extend beyond this we leave as quite an open question.  As a matter of fact we cannot assert that the internal processes of life follow the same laws as the motions of inanimate bodies; nor can we assert that they follow different laws.  According to appearance and general opinion there seems to be a fundamental difference.  And the same feeling which impels us to exclude from the mechanics of the inanimate world as foreign every indication of an intention, of a sensation, of pleasure and pain,---this same feeling makes us unwilling to deprive our image of the animate world of these richer and more varied conceptions.  Hertz, \eightit Die Prinzipien der Mechanik\eightrm, Leipzig, 1894 p.\ 45.}
allowed that one had to still keep an open 
mind between dualism and monism.  If the choice has to be made between 
monism \it plus\rm\ verificationism \it plus\rm abandonment of the axiomatic method, on 
the one hand, 
and dualism, on the other, then that would be an important fact to be established.  The reader 
is asked to proceed without assuming that logical positivism and verificationism 
are proved.  We adopt either Wittgenstein's definition of `meaning', \it viz\rm.,
the meaning of a statement is the state of affairs which would be the case \it if
\rm\ the statement were true, or any other normal meaning that would be accepted 
by Newton, Maxwell, Hertz, or the classical epistemology of Classical Physics.
The reader is asked to put aside any prejudices against counter-factual 
conditionals.  The experience of physics pedagogy in that we sometimes study 
two systems as if they were isolated, and then as if they were coupled, suggests 
that counter-factual conditionals are meaningful.  Perhaps the reason outlandish
philosophies (such as quantum logic, verificationism, restrictive positivist-style
rules as to what is meaningful and what is not, complementarity) have been resorted 
to has been the assumption that this axiomatic mess cannot be cleaned up.  Let us see.

Physicists of the twentieth century already cleaned up the physical mess of 
physical theory existing in Hilbert's day, and without introducing any new 
messes.  With this axiomatic mess cleaned up along Hilbertian lines, it is 
possible that Hilbert's Sixth Problem is essentially solved, provided that 
future developments in Physics do not introduce essentially new axiomatic 
messes.  But it transpires that it is necessary to abandon monism.

\centerline {\bf The Analysis of the Axiomatic Difficulty}

      Wigner$^1$ wrote several fundamental analyses of the problem of quantum measurement, and 
the formulation of the problem which we adopt here is due to him, he called it the problem of 
quantum duality.  He was aware of the style of thinking in Hilbert's orbit.  The deterministic axioms of 
Quantum Mechanics, in Weyl's formulation, assume, to begin with, that every 
system is a closed system.  To every system is associated a Hilbert space ${\Cal H}$, and a 
Hamiltonian operator $H$.  Every physical state of the system is described by a ray 
in that Hilbert space, as usual, we fudge the identifications and consider a non-zero wave 
function $\psi$ instead of the ray.  If the state of the system at time $t=0$ is $\psi_0$, 
then its state at time $t$ is given by 
          $$\psi_t=e^{{-2\pi itH}\over h} \cdot \psi_0 \in {\Cal H}.$$

     But there are separate, probabilistic axioms for measurement processes.  For every 
observable, there is a self-adjoint operator $Q$ on ${\Cal H}$.  If the system in the state 
$\psi$ undergoes a measurement process associated to this observable, the only possible 
results are the eigenvalues $\{\lambda_i\}$ of $Q$.  Assuming that $\psi$ is normalised 
and that its Fourier decomposition with respect to the normalised eigenvectors of $Q$ is 
           $\psi=\sum_i c_i v_i $, 
then the probability that the result will be $\lambda_i$ is $\vert c_i\vert^2$.  
Dirac added the following axiom.
If the result is $\lambda_i$, then the system is, as a result of the measurement process,
in the state given by the wave function $v_i$ even though such a transition or jump (called 
the reduction of the wave packet)  
disobeys Schr\"odinger's equation.
(We reserve discussion of this for a projected sequel.)

     On the other hand, the measurement apparatus itself must be a quantum system, possessed 
of a Hilbert space ${{\Cal H}}_n$ to describe its physical states ($n$ is 
the number of particles in the apparatus) and a Hamiltonian $H_n$ to govern its time-evolution 
if it were in isolation.  Then the state space of the combined system of the microscopic, as 
we will call it, system originally under discussion and which is being measured, and the 
macroscopic, as we will call it, measuring apparatus, is ${\Cal H}_n^{{\text com}}={\Cal H} \otimes {\Cal H}_n$ 
and the joint Hamiltonian is the sum of the two Hamiltonians which would have governed each 
system in isolation plus an interaction term
(this is almost tautological)
         $$H_n^{{\text com}}=H\otimes I_n + I\otimes H_n + H_n^{{\text int}}.$$

The problem of quantum duality, Wigner$^1$, was that we have two rather different 
mathematical descriptions of the same physical process and it is not at all clear how to compare 
them.  In the former description, we are implicitly treating the measurement apparatus as if it 
were a classical system which did not obey the superposition principle so that we are sure that 
the result of the measurement process is always a definite pointer position, as it is called, 
a macroscopic pointer visibly and definitely pointing to one or another of the various possibilities 
$\lambda_i.$  Bohr (as quoted in J. Jauch, E. Wigner and M. Yanase$^2$) always insisted that the measurement apparatus had to be classically describable
and classical in nature.   Heisenberg (cf.\ J. Bell$^3$) always insisted that we have to put a cut somewhere, marking 
off when we use the first three axioms to analyse things, and when we use the measurement axioms.
 Von Neumann (as reported in Wigner$^1$) showed that as long as we do put the cut somewhere eventually, it makes no difference 
where we put it.  Wigner contributed to this analysis, emphasising that on the classical side of 
the cut will always be the observer's consciousness, at least.  So the methodology of introducing 
a cut used to be considered the solution to the problem of quantum duality.
But this no longer commands a consensus in light of advances in mesoscopic engineering, detection of quantum mesoscopic chaos, macroscopic superpositions of states, and so on.  

     Einstein asked the vague but profound question whether or not the probabilities of the latter 
three axioms did not arise from some underlying deterministic dynamics in an analogous way to 
the way they did in classical statistical mechanics.  He seems to have thought that this would 
mean either revising Schr\"odinger's equation or perhaps introducing hidden variables.  But the 
logically sophisticated treatment of classical statistical mechanics due to C. Darwin and R. Fowler$^4$, and 
extended by A. Khintchine$^5$, does not introduce hidden variables.  For us, the method by 
which G. Ford, M. Kac and P. Mazur$^6$ carry out the Gibbs program for the case of harmonic oscillators with a 
cyclic nearest neighbour interaction is paradigmatic.  (J. Lewis and H. Maassen$^7$ has extended this.)  By focussing Einstein's question on 
Wigner's formulation of the problem, we can answer it positively, taking Schr\"odinger's equation 
as the underlying deterministic dynamics.  That is, we derive the probabilistic axioms 
from the deterministic axioms.   We make no other assumptions$^8$ (Farhi, Goldstone and Gutmann made several tacit \it ad hoc\rm\ assumptions), except that we have to 
introduce some sort of dictionary that compares quantum states with macroscopic pointer positions,
or else the comparison of the dual descriptions is logically impossible.
Basically, we push the cut out to infinity.  No real quantum system is exactly a measurement apparatus,
but the thermodynamic limit of quantum amplifying apparati becomes a classically describable 
measurement apparatus which exactly verifies the three probabilistic axioms in the limit as $n\rightarrow
\infty$.

     Physically, this model has much in common with aspects of previous work of 
H. Green$^9$ and A. Daneri, A. Loinger, and G. Prosperi$^{10}$.  
One important physical difference with Coleman---Hepp$^{11}$ is that the notion 
of macroscopic which we will introduce is the opposite of their notion of local observable
(which is from the theory of infinite volume thermodynamic limits, but not physically
appropriate here).  
The idea that coupling a Brownian mote to a negative 
temperature amplifier will amplify the motion of the 
mote from quantum motion, where observables do not 
commute, to classical motion, where the non-commutation 
becomes negligible, is perhaps due to J. Schwinger$^{12}$.  
Our model is that of a non-demolition measurement, 
however, and we assume that the amplifier exerts zero 
force on the incident particle.  

     This has mathematical similarities to the literature on the problem which takes the open system
approach.  But the physics is different.  We make the transition to irreversible classical stochastic 
dynamics a function of the coupling between the apparatus and the microscopic system.  
W. Zurek$^{12}$ and others$^{14}$ such as M. Collett, G. Milburn, and D. Walls$^{15}$ use a sort of open 
systems approach in that they put the interaction which 
is supposed to turn quantum amplitudes into classical probabilities in the coupling with 
the environment instead of in the coupling with the amplifier.  
We assume instead a closed joint system.  
 In principle, this difference should be detectable by experiment.
We also predict that the degree of validity of the probabilistic axioms should worsen as the size 
of the amplifying apparatus approaches the mesoscopic or even microscopic.  This should be 
detectable by experiment as well.  
The only real novelty is the notion of macroscopic which we introduce.  The need for a precise 
definition of macroscopic has long been felt.  Indeed, it is not possible to compare the two dual 
descriptions of the measurement process unless some sort of dictionary is provided.  Von Neumann 
and Wigner$^{16}$ (footnote 203) seems to have missed the need for a relatively sophisticated dictionary, and simply 
assumed that a macroscopic pointer position 
corresponded, more or less approximately in the strong topology on ${\Cal H}_n$, with a large set 
of wave functions.  It seems to be this unquestioned assumption which has been the obstacle to 
progress along Wigner's original lines.  Abandoning it may mean the abandonment of the ancient dream 
of psycho-physical parallelism.   ([16], p. 223.) 

     In connection with this I quote P. Dirac$^{17}$.  ``And, I think it might turn out that ultimately Einstein will prove to be right, \dots that it is quite likely that at some future time we may get an improved quantum mechanics in which there will be a return to determinism and which will, therefore, justify the Einstein point of view.  
But such a return to deteminism could only be made at the expense of giving up some other basic idea which we now asume without question.  We would have to pay for it in some way which we cannot yet [1977] guess at, if we are to re-introduce determinism.''  
In this regard I wish to point out that there is almost no credible evidence in favour of the hoary 
psycho-physical parallelism, and there is a good deal of evidence for the time-dependent form of 
Schr\"odinger's equation's being absolutely linear and unitary.  

     It will be noticed that our physical interpretation, then, is that there are only waves.  There are 
no particles.  The problem of quantum duality, then, becomes the problem of calculating, relatively 
explicitly, how it is that the interaction of the wave of, say, a microscopic system with one degree 
of freedom (called, sentimentally, an incident particle) interacts with the wave of the amplifiying 
apparatus to sometimes (with a definite probability) produce a ``particle-event,'' i.e., the registering 
of a loud click on the part of tha apparatus, and other times, produce no such detection event.  Thus the 
seeming particle-events, the seeming detection of particles, is an artifact of the amplification process 
in the interaction of two waves.  
     
     Feynman, famously,$^{18}$ did not think so, but he did think that perhaps a little more could be said 
about the problem of wave-particle duality (p. 22): 

{\narrower\baselineskip=12pt\smallskip``We and our measuring instruments are part of nature and so are, in principle, described by an amplitude function satisfying a deterministic equation.  
Why can we only predict the probability that a given experiment will lead to a definite result?
From what does the uncertainty arise?  Almost without a doubt it arises from the need to amplify the effects of single atomic events to such a level that they may be readily observed by large systems. \smallskip}  

{\narrower\baselineskip=12pt\hskip-10pt ``\dots In what way is only the probability of a future event accessible to us, whereas the certainty of a past event can often apparently be asserted?\dots Obviously, we are again involved in the consequences of the large size of ouselves and of our measuring equipment.  
The usual separation of observer and observed which is now needed in analyzing measurements in quantum mechanics should not really be necessary, or at least should be even more thoroughly analyzed.  
What seems to be needed is the statistical mechanics of amplifying apparatus. \smallskip}

It is usually not necessary to think about the philosophical meaning of probability, as A. Sudbery$^{19}$ 
remarks, 
``It is not in fact possible to give a full definition of probability in elementary physical terms.''
``Attempts to define probability more explicitly than this are usually either circular \dots or mysterious \dots This is not to say that the question of what probability means, or ought to mean, is not interesting and important; but the answer to that question, if there is one, will not affect the properties of probability that are set out here, and we can proceed without examining the concept any further.''
All that is necessary is to straightforwardly imitate the classical methods of statistical 
mechanics, in a physically appropriate way.  Remarkably, the result we arrive at corresponds exactly to J. von Plato's philosophically motivated amendment$^{20}$ of the traditional frequency theory of probability.  The naive frequency theory suffered from grave logical circularities and Professor von Plato fixed these by incorporating a physical, ergodic dynamics.
mechanics.  His work has not received the attention it deserves because he had to rely on 
the underlying dynamics' being deterministic, so it was usually assumed that his purely 
logical, almost antiquarian concerns, could not be relevant to a quantum world.  This turns out to be a  misapprehension.  It is, however, useful to revise Professor von Plato's theory in 
two ways.  Firstly, time averages are not, for us, the definition of probability, they 
are the definition of measurement.  Secondly, we need to extend it to dynamical systems 
which although not ergodic, are, because of their large number of degrees of freedom, 
approximately ergodic in some respects, as envisioned by Khintchine.

\centerline{\bf Analysis of the Role of Probability}

\vskip-8pt
Each axiom by itself would be a viable candidate.  It is only when all 
five (let alone all six) are taken together as a system that we get a mess.
They do not form an axiom system in the usual sense of the word.  This aspect 
was analysed in the previous section.  The aspect to which we now turn is 
the role and status of the concept of (physical) probability.  From now 
on the word probability by itself always refers to physical probability, and 
never to mathematical probability.  The operationalisation of the concept of 
probability is going to be retained unchanged as unproblematic: the great 
practical successes of Physics demand this: we always operationalise the 
process of measuring a probability as counting the number of successes in 
trials (repeated trials of an experiment) and dividing by the number of 
trials.  The fact that we always get different answers is not unusual for 
practical physics, this was true in astronomy as well even in ancient times.
But since we do not adopt positivism, this \it operationalisation\rm\ does not 
have to be accepted as the \it meaning\rm\ of the concept of probability.

If we are to formalise the language of Physics, to construct a formal 
axiomatic system with any such axioms and concepts in it, we must pick which 
concepts are primitive (undefined), state which axioms they satisfy, and 
define all other concepts in terms of the given ones.  Derive all future 
physical propositions as theorems, following from the definitions and axioms 
with no new assumptions.  Hilbert would have asked us to investigate 
whether probability is primitive or not, and, if primitive, what are its 
axioms, and if not, what is its definition in terms of system, state, 
and Hamiltonian?  Likewise for measurement.

\centerline{\it The logical problem of circular definition: 
an existing consensus }

The usual `definition' of probability 
is well-known to be logically imprecise, even circular.

`If the experiment is repeated a large number of times it will be found that each 
particular result will be obtained a definite fraction of the total number of 
times, so that one can say there is a definite probability of its being obtained 
any time the experiment is performed.'\plainfootnote*{\baselineskip=8pt \eightrm P.\ 
Dirac, \eightit The Principles of Quantum Mechanics\eightrm, Oxford 1930, p.\ 10.}

Now, it is not carping or nitpicking to point out that this is, if taken 
\it au pied de la lettre\rm, untrue.  And, if not so taken, it does not count 
as a definition, from the Hilbertian standpoint.  Dirac has not succeeded in 
defining the concept of physical probability.
It is not strictly true because if one large number is a prime, and another 
large number is a different prime, it is impossible for the same definite 
fraction of successes to be obtained.

Features which do not, to my mind, count as defects in an \it operationalisation\rm,
can be fatal flaws in a logical definition.  And that is what we have here.

Many famous mathematicians have shared substantially the same understanding of 
the logical flaw, circularity, in any attempt to define the concept of physical 
probability that remains close to the idea of frequency.  For example, Burnside,\plainfootnote
*{\baselineskip=8pt \eightrm W.\
Burnside, ``On the Idea of Frequency,'' \eightit Proc. Camb. Phil. Soc., \eightbf 22 \eightrm(1925), 726.}
\ in a polemic with Fisher, and also 
Littlewood and Kolmogoroff whom we will quote.

Firstly, Littlewood\plainfootnote\dag{\baselineskip=8pt \eightrm
Littlewood, \eightit A Mathematician's Miscellany\eightrm, London, 1956, p.\ 32.} explains, in his insular way, the same point we have made 
in distinguishing between the task of defining the mathematical concept of 
probability and the physical concept, and also between the tasks of 
asserting mathematical propositions about mathematical probability, and 
scientific propositions (or axioms) about physical probability.

\baselineskip=12pt \eightrm 

{\narrower
`Mathematics \dots has no grip on the real world;
if probability is to deal with the real world it must contain elements 
outside mathematics, the \it meaning\eightrm\ of `probability' must relate 
to the real world; and there must be one or more `primitive' propositions
about the real world, from which we can then proceed deductively (\it i.e.
\eightrm\ mathematically).  
We will suppose 
(as we may by lumping several primitive propositions together) 
that there is just one primitive proposition, 
the `probability axiom', and 
we will call it `\it A\eightrm' for short.\smallskip}

{\narrower`\dots the `real' probability problem; 
what are the axiom \it A\eightrm\ and the meaning of `probability' to 
be, and how can we justify \it A\eightrm? 
It will be instructive to consider the attempt called the `frequency theory'.  It is 
natural to believe that if (with the natural reservations) an act like throwing 
a die is repeated $n$ times the proportion of 6's will, \it with certainty\eightrm, 
tend to a limit, $p$ say, as $n \rightarrow \infty$.  (Attempts are made to 
sublimate the limit into some Pickwickian sense---`limit' in inverted commas.  
But either you \it mean\eightrm the ordinary limit, or else you have the problem 
of explaing how `limit' behaves, and you are no further.  You do not make an 
illegitimate conception legitimate by putting it into inverted commas.)  If we 
take this proposition as `\it A\eightrm' we can at least settle off-hand the other 
problem, of the \it meaning\eightrm\ of probability,  we can define its measure for 
the event in question to be the number $p$.  But for the rest this \it A\eightrm\ 
takes us nowhere.  
\dots Now an \it A\eightrm\ cannot assert a \it certainty\eightrm\ about a 
particular number $n$ of throws, such as `the proportion of 6's will \it 
certainly\eightrm\ be withing $p\pm\epsilon$ for large enough $n$ (the largeness 
depending on $\epsilon$)'.  
It can only say 'the proportion will lie between $p\pm\epsilon$  \it with at 
least such and such probability (depending on $\epsilon$ and $n_o$) whenever 
$n>n_o$'\eightrm.  The vicious circle is apparent.  We have not merely failed to 
\it justify\eightrm\ a workable 
\it A\eightrm; we have failed even to \it state\eightrm\ one which would work if its 
truth were granted.\smallskip}  


\rm \baselineskip=24pt
Essentially the same criticism was made by
Kolmogoroff in his contributed chapter to Alexandroff, Kolmogoroff, and Lavrentieff,
ed.s, {\it Mathematics its content, Methods, and Meaning}, 2nd ed., Moscow,
1956, we cite from a Cold War translation published in 1963 in Cambridge,
Mass., p.\ 239, ``\dots it is clear that this procedure will never allow us to be
free of the necessity, at the last stage, of referring to probabilities in the
primitive imprecise sense of this term.'' 

\centerline{\it Three approaches in the literature}

Many naive attempts to fix this problem have been published, but none have 
succeeded.  (Professor von Plato's is far from naive and
is postponed until later.)  We will 
only deal with von Mises, Hardy,
and Farhi--Goldstone--Guttmann.

Richard von Mises carefully analysed the circularity involved in trying to 
say anything like, the larger the number of trials, the greater will be the 
likelihood that the ratio of successes to trials will be, for all observers, 
within a small epsilon of each other, \it etc\rm.  This word `likelihood' 
is either a synonym, in which case this attempt is circular, or needs a 
definition, and none has been provided.  Furthermore, von Mises opined that 
no rigourous foundation could be provided for Statistical Mechanics and 
so one must abandon anything like the frequency theory and take another 
road.  His suggestions for a positive solution need not detain us, even 
though related ones were proposed by Alonzo Church, later, and in the 60's, 
Kolmogoroff, since they are remote from Statistical Mechanics, unphysical, 
and have not been able to justify the use in Physics of probability the 
way Hilbert demanded.  His opinion that a rigourous 
justification was impossible has been disproved, by recent successes in 
the theory of Hamiltonian Heat Baths, stemming from the breakthrough work 
of Ford--Kac--Mazur$^6$ in 1965.

Next we will consider one of the Quantum Information style axiomatisations, 
this one due to Lucien Hardy.

{\baselineskip=12pt\narrower\smallskip\noindent `Axiom 1 Probabilities.  Relative
frequencies (measured by taking the proportion of times a particular outcome is observed)
tend to the same value (which we call the probability) for any case where a given
measurement is performed on an ensemble of $n$ systems prepared by some given
preparation in the limit as $n$ becomes infinite.
\plainfootnote*{\baselineskip=8pt \eightrm\hskip-3pt L.\ Hardy, \eightit Quantum Theory from Five Reasonable Axioms\eightrm\ arxiv.org/quant-ph/0101012.}
\smallskip}
Here, probability is defined (but `result' and `measurement' are primitive).
The words `any case' are subtly vague, as we shall see.

Could Hardy's suggestion as to `the limit of this ratio as the number of 
experiments increases to infinity' be adopted? Let us concede the meaningfulness 
of an actual or potential infinity of repeated experiments.  The missed opportunity 
here  
is, what is the definition of `limit'? In mathematics, there is no one canonical 
definition of limit.  There are different ones which apply to different situations.
There is the limit of a sequence, of values of a function, the limit in mean of 
a sequence of functions, weak limits, \it etc\rm.  (In this paper, we will 
define a new kind of thermodynamic limit, and argue that it is just what is 
needed to fix this problem.)
None of these apply unless we can argue that the physical situation of 
repeated experiments is modelled by the particular mathematical concept 
chosen.  (We will, eventually, argue that the axioms allow us to 
model measurement, but not other repetitious situations, by the Hamiltonian of an amplifier 
with an extremely large number of degrees of freedom, but a given, fixed, 
invariable initial condition, which we will analyse 
the way Jeans suggested).

It is clear from context that Hardy intends the elementary notion of the 
limit of a sequence.  So he has defined probability incidentally and by the 
way, and the major empirical content of his axiom is that the limits 
always exist.  This is surprisingly problematic (no surprise to the experts).

Firstly, mathematical sequences are deterministic: timeless, they have 
no dynamics.  The one-millionth result is already determined in advance.  It 
would be intuitively and emotionally odd if this was the right concept to 
use for Hardy's purposes.  But let that pass.  Secondly, and more importantly,
not all sequences possess the property he demands, that the fraction does indeed 
have a limit.  This is why there is significant empirical content to his axiom:
only those sequences are physically possible which do indeed 
have a limit.  

This simply does not work.  If one experimental situation 
has one sequence, $f_n$, whose limit is one-half, and a different one has a different 
sequence, $g_n$, say, with some limit, too, (for simplicity, we may assume it is one-half), 
clearly we may physically construct an electronic AND-gate or what-not and 
make logical, Boolean combinations of the results, and the resulting sequence 
will be equally physical.  But it is easy to construct such sequences whose 
logical combinations do not possess any limit at all.
So we see that the words `any case' are not really well-defined.
It seems pointless to try to introduce a countable hierarchy of patches 
to this theory by postulating that every reasonably physically implementable 
combination of sequences must also possess limiting frequencies\dots
(Perhaps this is the same problem von Mises ran into with his theory of 
selections from collectivities?)
There was never anything physical about Hardy's postulate to begin with.

\vskip-8pt
The above point is an immanent but physical critique of the logical consequences 
of his empirical assumption.  In my opinion it is decisive.  
An easier (but not decisive) point can be made: the usual laws 
of probability do not in fact guarantee that such limits exist, and the 
propensity theory does not make this assumption.  
The laws of probability clearly state that the probability that in two rival infinite 
runs of the same experimental setup, two different limiting ratios will occur, or 
one limiting ratio will occur in one but no limit in the other, or, no limit in 
either, is zero, but probability zero does not mean impossible, such a `violation' of our expectations is physically 
and logically possible. So Hardy is 
(and he is allowed to do this, but it seems odd and unwise) explicitly postulating 
that physics does not obey the laws of mathematical probability, that physical 
probability, as he has defined it, does not satisfy the axioms of mathematical 
probability (and this carries over, later, to the laws of non-Kolmogoroffian 
probability in his other more developed contexts).  (His postulate wreaks 
havoc with the mathematical notion of stochastic independence.)

In brief, the usual interpretation of the laws of mathematical probability as 
applied to the fair toin coss situation is that it is physically and logically 
possible for every single toss to come up heads, and that such a `violation'
does not contradict 
the statement that the probability of a result of heads is one-half.  Hardy wishes 
to make the frequency theory more important than the mathematical laws.  To 
accomplish this axiomatically, he would need to reconstruct the mathematical 
theory of probability so that it would match his physical postulate. 

Can the much-cited paper of Farhi--Goldstone--Gutmann  be taken 
at face value?  Its argument rests, crucially, on a careless fallacy of equivocation.  They 
use the same notation for two very different meanings, and at a crucial point 
in their argument, switch, impermissibly, betweeen the two meanings.  I do not 
doubt for a minute that there could be some physical motivation for supposing 
that there is a statistical correlation, in all normal applications, between 
the two different meanings, but they neither make such an assertion explicitly, 
nor would such an unproved asserion be useful in an axiomatic analysis.  Indeed, 
their explicit claim to have reduced their result to the principle of invariance 
of physical results under change of co-ordinates suggests that they made the 
switch inadvertently.

Their key claim is that a formulation of quantum mechanics without explicit 
reference to probabilities allows them to instead appeal to the general philosophical 
principle that physical results must be independent of the choice of 
coordinates used to calculate them,  the appeal occurring at a point where
a certain map, which they 
construct, would, they claim, have to be an isometry for this principle to hold good.  
By construction, that it is an isometry means a probability measure 
is defined uniquely.
 This might be fair if they had in 
fact defined the map and the measure in the very places they claim to have defined them,
but they have not.

More precisely,  
they alert us to their desire ''to contruct a transformation from the basis \dots 
of normalised states to the basis \dots of infinite norm states.''
(This is evidently in order to later appeal to the general philosophical principle 
as stated much previously in their paper, 
`the necessity to describe physics in a basis 
independent manner.')
This would be (not a `transformation', but a) change of basis matrix if it were not for the fact
that the two `bases' are of different cardinalities, one countable, 
and the other un-countable. What they must mean, from context, is that 
they desire to exhibit two formulae, (TII) will express the co-ordinates of 
an element $\Psi$ with respect to the countable basis in terms of any given 
co-ordinates with respect to the uncountable basis, and (TIII) will be the 
other way round.  That is, the map is the identity map on a quite concrete 
Hilbert space, $V_c^{(\infty)}$ the element $\Psi$ does not change.

\vskip-8pt
An `infinite norm state', as usual, is not an element of the Hilbert space 
$V_c^{(\infty)}$ itself, but is a linear functional defined on a dense subspace.
The formulas are, then, only valid on a dense subspace.  This, probably, is 
why they equivocate between whether, for a fixed infinite sequence $\{j_n\}$, 
their `infinite norm state' $\langle b;\{j_n\} \vert$ is a linear functional,
or whether it is a mere formal symbol which, when applied to a vector $\vert \Psi\rangle
\in V_c^{(\infty)}$, yields $\langle b;\{j_n\} \vert \Psi \rangle$, an $L^2$-equivalence
class of functions on the measure space $Y$ of all infinite sequences $\{j_n\}$. 
In the former case, we are dealing with alternate co-ordinate descriptions of 
the same vector.  (But only on a dense subspace).  In the latter case, we are 
dealing with a concrete map between two different spaces which can be extended 
to be an isometry.  That these two different set-ups can be described by 
formulas which look alike is irrelevant to general philosophical principles!
This is not simply a change of basis.

As we have just elucidated, they switch from interpreting this `basis independence'
as alternate descriptions of the same vector, to interpreting it as a map between 
two different vector spaces which must be extendable to an isometry on their 
respective completions.  This is fallacious.  Our analysis of every naive 
version of the frequency theory shows that the sticking point is what to do about 
this negligible, but physically possible, set of `violators', of measure zero.  What they 
have done is defined a physically meaningful operator $F$ on only a dense 
subspace of the Hilbert space, then, mapped this by an isometry to a space 
where they can extend it by a neat formula.  But on this other space, it no longer 
has the same logical and physical significance, when extended.  The operator 
$F$, when extended, no longer is a `frequency' operator.

\vskip-8pt
In fact, the paper is just a tangle of conceptual confusions (from the 
point of view of axiomatic investigation, or from 
the standpoint of Hilbert's Sixth Problem).

\vskip-8pt
On page 370 they correctly assert just what we mentioned in our discussion of 
Lucien Hardy, that the strong law of large numbers 
says that the set of all sequences whose proportion of heads does not 
settle down to any limit or settles down to the wrong limit, has measure zero.
Then they immediately pass to a non-equivalent statement: $\vert\psi\rangle 
^\infty$ is an eigenstate of the operator $F(\theta_i)$.  They immediately 
assert that this last statement is a `quantum version' of their previous 
statement.  This is just 
confused.  Their axiom \bf PIV'\rm\ clearly states that if a quantum system 
is described by the state $\vert \theta_i\rangle$ then a measurement of 
the observable `will yield the value $\theta_i$'.  They just admitted that 
there is a non-empty set of sequences for which this is not true.  This is 
Pickwickian, in the Littlewoodian sense, applied to `will yield'. Whatever 
can they mean, formally and axiomatically, by `will yield' if in fact they 
admit\baselineskip=8pt \plainfootnote*
{\baselineskip=8pt \eightrm
`For a recent expression of the view that on the contrary there is no real 
problem, only a `pseudoproblem', see J.\ M.\ Jauch, \eightit Helv. Phys. Acta \eightbf37, 
\eightrm293 (1964).  \dots current interest in such questions is small.  The 
typical physicist feels that they have long been answered, and that he will 
fully understand just how if ever he can spare twenty minutes to think about it.'
J. Bell and M.\ Nauenberg, ``The Moral Aspect of Quantum Mechanics,''
in  De Shalit, Feshbach, and de Hove, eds., \eightit Preludes in Theoretical Physics, 
in Honour of Weisskopf\eightrm, Amsterdam, 1966, 279-86}
\baselineskip=24pt  that it might not yield?

In general, the difference between the strong law of large numbers and the 
weak law of large numbers is irrelevant for addressing this problem.  If the 
probability is continuous, probability zero does not mean impossible, 
so no advance has been made over saying the event of a violation of the 
long-run average's settling down to a definite fraction is `unlikely'.
(Kolmogoroff was 
certainly an expert in the difference between the strong law and the weak law 
that Farhi \it et al\rm. make such a big deal of, and his considered opinion on 
the matter has been recorded.)
One could introduce a new primitive physical term, `negligible', and 
add a new postulate, probability zero means the event, although possible, 
is negligible. Let anyone who wishes to rely on this, be 
conscientious about writing down all the details.
Then the mathematical notion of passing from one function to another 
one in its $L^2$-equivalence class could validly be the model of the 
physical notion of neglecting negligible differences.

Lastly, they still assume observable and measurement as primitives, without 
any attempt to elucidate the relationship between physical measurement 
and other physical processes.  This means they have proved uniqueness 
but not existence.

It would be unlikely if the solution to the problem of quantum measurement 
did not rely on the physics of measurement, but only on axiomatic analysis.

\centerline{\it A new approach}

The structure of the axioms themselves suggest that there are two scales going 
on, and that probability is a multi-scale phenomenon.  Feynman, as quoted, went 
into this physically.  That will be our approach.  Then, to be consistent, 
one must make measurement a two-scale phenomenon.  Further, since observables 
only crop up in the `large' axioms, this strongly hints that they are part 
and parcel of the concepts to be derived, rather than assumed.  
Our approach will be to use the usual 
Hamilton-Liouville abelian dynamical variables, and we will not refer to them as 
observables. We will derive the new properties of 
(quantum, non-commutative) observables from the definition 
of measurement and probability.  This will become clearer when the model is 
concretely spelled out.  This marks the major logical change between this 
paper and all previous papers, even recent ones such as Allahverdyan--Balian--
Nieuwenhuizen, which assume the usual properties of observables even while 
they are deducing them from the Hamiltonian and Schroedinger's equation.

Mere logical combinations of concepts are always admissible since they do 
not rely on new physical concepts.  Since the mathematical theory of probability 
shows that it is equivalent to give the probability measure, or to give 
the mathematical expectation of every integrable function (since this allows 
the measure to be re-constructed from these expectations), we need only give 
a physical definition of the expectations.  Then the same logical processes 
that derive mathematical probabilities from the expectations, can be used 
to derive the numbers representing the physical probabilities from the numbers 
representing the physical expectations.  And that is how we will proceed.

Einstein's insight, that quantum mechanical probabilities might be a 
statistical mechanical phenomenon, suggests we look carefully at what Hilbert 
asked for in the derivation, within Classical Mechanics, of probabilities 
and irreversible phenomena, from deterministic equations of reversible evolution.
To do this we must not postulate ignorance or probability distributions, and 
must, still, define probability.  Now Professor von Plato has done this if the dynamics 
is ergodic.  Also, Darwin--Fowler set up Statistical Mechanics without the use 
of probability, they used a neutral term `weight' to make it clear that 
they were merely making a mathematical definition.  In the retrospect afforded 
by the work of Ford--Kac--Mazur, and looking forward to use in quantum 
measurement, it is possible to assemble the different contributions of 
Darwin--Fowler and Professor von Plato along lines Khintchine conjectured, at least 
in some special cases, to give an answer Hilbert would have accepted as 
logically `clean.'  We will do this in the next section, and fortunately it 
applies to at least a toy model of a quantum amplifier, which will follow.

Although logically and axiomatically `clean', it is convoluted, just as was 
Russell's theory of description.  But its empirical content is still very 
close to the logically muddled frequency theory, and this is a serious 
recommendation.  That it is logically convoluted is by no means a piece of 
evidence against it, since, as Russell remarked in the Introduction to the \it Principia\rm, 

{\narrower\smallskip \baselineskip=12pt `The grammatical structure of language is adapted to a wide variety of 
usages.\dots  Language can represent complex ideas more simply [than 
simple abstract ones].  The proposition ``a whale is big'' represents 
language at its best, 
giving terse expression to a complicated fact,
while the true analysis of ``one is a number''
leads, in language, to an intolerable prolixity.' \smallskip}

Russell, according to Wittgenstein, first showed us that the logical 
structure of a proposition need not be its apparent grammatical structure.
And he had the theory of description in mind.  The simple word `the' 
required an elaborate `unpacking' to get at its logical function as part 
of a description.  It should not be surprising that if, on physical 
grounds, the concept of probability should be a two-scale process, 
then too, its logical structure should be very far from the apparent 
grammatical structure of the frequency theory or its use in Physics.
I hope the reader will excuse any consequent intolerable prolixity.

\centerline{\bf The Logical Structure of Statistical Mechanics}

Our procedure is modelled on that of Ford--Kac--Mazur, which indeed is modelled on the 
usual understanding of the Gibbs program.  
(Later we will point out two significant alterations, inspired by work and far-seeing 
remarks of Wiener and Khintchine.)

In Sir James Jeans, the very sane idea is expressed that Statistical Mechanics 
is the study of the statistical properties of the normal state\plainfootnote
* {\baselineskip=8pt \eightrm`The first 
object of statistical mechanics is to determine all the ``normal'' properties of 
such an assembly and correlate them with the properties of matter in bulk as we 
know it, when the assembly is in complete equilibrium. \dots We may define the 
``normal'' properties of the equilibrium theory in the manner of Jeans, or perhaps 
more naturally as \eightit all those properties which the assembly possesses on a 
time average\eightrm.' 
Fowler, \eightit Statistical Mechanics\eightrm, Cambridge, 1929, p.\ 8.}
 (or trajectory, 
same thing).  The statistical properties are such as mean, dispersion, correlation, 
auto-correlation, three-point correlations, and so on.  None of these ideas 
depend on the concept of probability, they are `statistical' in the precise 
sense of being summary statistics meant to convey some idea of the properties of 
a large (even infinite) amount of data.  The data are conceived of as quite 
deterministic.  The definition of normal\baselineskip=8pt\plainfootnote*{\baselineskip=8pt \eightrm Ehrenfest and Ehrenfest,
\S 18b, \eightit The Conceptual Foundations of the Statistical Approach in Mechanics, \eightrm p.\
41
} \baselineskip=24pt
 is not given with full logical rigour, 
but this will be fixed later.  Jeans's definition of normal\baselineskip=8pt\plainfootnote
{\dag}
{\baselineskip=8pt \eightrm Jeans, \eightit op.\ cit.\eightrm, 
  Ch.\ III, \S\S 52,58, and Ch.\ V,
\S\S 87.} \baselineskip=24pt
 is, a state is normal 
if the set of other states, all of whose statistical properties are the same as 
this given state, is infinitely probable.  By this he means that as the number 
of degrees of freedom of the system increase, the ratio of the Liouville measure 
of this set of states, divided by the measure of all other states which the system 
could pass into, goes to infinity.  We see at once that no real idea of physical 
probability is being relied on here, just Liouville measure (or, rather, the 
measure inherited on the surface of constant energy from Liouville measure, or, 
if more controllable integrals of the motion are present, the smaller dimensional 
surface of `accessible states').

Now every such statistical property involves some kind of time average.
$$ \lim_{T\rightarrow \infty } \frac1T \int_0^T f(x_t) dt.$$  The 
Gibbs procedure was to first argue that these time averages would be equal 
to phase averages 
$$ \int_{\Cal M_E} f(x) d\mu(x)$$
with respect to the microcanonical distribution, and also 
with respect to the canonical (Gibbs) distribution.  
$$ \int f(p,q) \alpha e^{-\beta E} dpdq \text {\quad where\ \ } \ \beta \text { \ is inverse temperature}.$$ 
The argument was not 
rigourous.  It has been made rigourous in enough special cases that we can 
have confidence that von Mises was wrong, it will be possible to give the 
procedures of Statistical Mechanics an axiomatically sound derivation as 
Hilbert asked for.

Consider a sequence of Hamiltonian (linear, conservative) dynamical systems
$M_n$ with degrees of freedom $n\rightarrow\infty$.
Each system should in some physically intuitive sense be the `same, except for 
the increase in degrees of freedom'.  
The 1965 breakthrough of Ford--Kac--Mazure was to derive Brownian motion 
(more precisely, the Ornstein--Uhlenbeck process) from a sequence of assemblies 
of harmonic oscillators with symmetric and cyclic couplings, increasing in 
strength, and a kind of renormalisation.

The procedure was to consider a dynamical variable 
$f_n$ on each finite system, $M_n$ (having Hamiltonian, say, 
$$H_n = \sum {p_i^2\over2} + \frac12 Q_n(q_o,\dots q_n) \text {\quad for }\ Q_n \text {\ a quadratic form }  .$$  It was the 
momentum $p_o$ of the Brownian mote. 
They then took a kind of phase average of this, its autocorrelation function, 
$g_n(\tau) = \int f_n(p,q) f_n(p_\tau, q_\tau) \alpha e^{-\beta E} dpdq $.  (It has since been shown that their results are insensitive to the 
choice of the mixed state of the heat bath.)  They then passed to the limit, $g_\infty$, as $n$ approaches $\infty$ 
using a cut-off procedure.  

I want to emphasize the following logical aspects of the procedure.
Each $M_n$ is first replaced by a Gaussian, deterministic, 
stationary, stochastic process (arising from the Gibbs distribution artifially imposed 
on it).  
None of these processes are ergodic or Markoffian, so they cannot be Brownian 
motion.  Next, each process is completely characterised by one datum, its 
phase auto-correlation function.\baselineskip=8pt\plainfootnote*
{\baselineskip=8pt \eightrm A standard theorem in stochastic processes says that a Gaussian, stationary stochastic process is determined up to stochastic equivalence by its phase auto-correlation function, and is Markoffian if and only if that function is exponential decay.  Further, all higher 
auto-correlations are easily calculable in terms of the two point ones referred to.}
\baselineskip=24pt So, each $M_n$ 
is replaced by this one function $g_n$.  The individual states are totally gone.
Now this sequence of functions has a limit, uniform on compact sets.  (Each $g_n$ 
is continuous and bounded and quasi-periodic).  The limit function is exponential 
decay.  We have, thus, first passed to a `dual object' (the $g_n$), then taken a 
limit.  Now we un-dualise.  There is a unique (up to stochastic equivalence) 
stochastic process which is Gaussian, stationary, and has that limit function $g_\infty
$ as its phase auto-correlation function.  It is not deterministic, and that is how 
probabilities arise from determinism if Statistical Mechanics is done logically 
carefully.
(Furthermore, K. Hannabuss's use$^{23}$ of the Sz.\ Nagy dilation is in much the same tradition: the dilation is 
constructed simply from the correlation functions.)

Notice the following important point.
Because we are interested in phenomena in the limit as $n$ approaches $\infty$, yet one cannot 
directly compare a value of $f_n$ on a vector $v_n$ with $f_{n+1}(v_n)$,  
we have to study statistical properties instead of individual points 
such as $f_n(v_n)$.  This is the logical reason for adopting such a procedure.

Because we have adopted such a procedure, we do not need to construct a limit dynamical 
system whose individual trajectories are composed of limits of the individual trajectories
of the finite systems.  But we adopted such a procedure because in fact it is logically 
impossible to do such a construction.  This is not a direct sort of limit.  The states 
of the limit dynamical system are not constructed from limits of trajectories or states 
of the finite systems.  It is a double-duality sort of construction.

Because Ford--Kac--Mazur present this as a `result' of imposing the Gibbs distribution early, 
it seems to most undergraduates that the probabilities arise from the imposition of 
this distribution, and much pedagogical ingenuity has been expended on motivating 
this particular distribution.  But the results are robust and do not much depend on the 
particular distribution (except that it should not be negative temperature!!!) so 
this is a red herring.

This does not fit into the Jeans--Darwin--Fowler scheme, but it is routine to adapt 
these results so that it does.  
(Although it does not seem to be in the literature anywhere.) 
 As $n$ increases, the measure (on a 
surface of constant energy) of states which are within $\epsilon$ of obeying the 
law of equipartition of energy and have nearly equally distributed plus and minus 
signs (`random', intuitively speaking) becomes overwhelming, becomes infinitely 
probable, in Jeans's terminology (as is well known).  Hence such states are the normal state because 
direct calculation shows their time autocorrelations are all approximately equal 
and tend, in the limit, to the same exponential decay typical of Markoffian processes.
And are all approximately equal to the phase autocorrelations calculated by 
Ford--Kac--Mazur explicitly.  (As always, the phase calculations were much easier 
than the time averages, and, furthermore, Ford--Kac--Mazur did not have to worry about normal 
state because the average wiped out the contribution of the `violators', as we have been calling them.)

This suggests that their postulation of a heat bath probability distribution is without 
real foundational significance, it is simply a crutch to circumvent a difficult 
calculation and a difficult argument.  But in this linear model, the difficulties  
are not great.

The limit stochastic dynamical system Ford--Kac--Mazur obtained, the Ornstein--Uhlenbeck 
process $X(\alpha,t)$ (for $\alpha\in [0,1]$ with Lebesgue measure) has a normal 
cell: almost all $\alpha$ are such that the sample path $X(\alpha,t)$ as $t$ 
varies, has time auto-correlation equal to the phase autocorrelation with respect 
to all $\alpha$ (which is exponential decay) (since it is ergodic) and so it is a limit dynamical system 
for the $M_n$ with respect to time auto-correlations, too.
As a `dynamical system' it is a little odd.  The `initial conditions' are the $\alpha$, which 
are not points on the `trajectories' (the sample paths).  We have, in fact, 
left the category of dynamical systems we were working in: the `limit object', if we 
insist on having one, is not a stochastic dynamical system in any usual sense, it 
is merely an abstract stochastic process constructed out of the correlation 
function.  We needn't really have undualised at all! (And, actually, Ford--Kac--Mazur 
did not trouble themselves to undualise, they simply exhibited the correlation 
function.)  As far as the philosophy of probability is concerned, there seems to 
be an important lesson in this: probability only arises if we actually depart from 
the category we were working in, to study an idealised approximation which really 
belongs to a quite different category.$^{26,28}$

This approach makes it much clearer that probabilities arise only in the limit 
object, and only because it is a double duality.  They did not arise through 
coarse-graining, subjectivity, ignorance, loss of information, or any extraneous 
new concept.  They are a merely linguistic feature of an unphysical limit which is 
a good approximation to certain limited, statistical, bulk aspects of $M_{10^{23}}$.

Therefore probability can be defined explicitly in terms of this sequence and the 
sequence of observables.

Khintchine, in the period from 1943 until his death in 1953, pointed out how 
true ergodicity ought not to be needed, and how the Darwin--Fowler logical 
outline (which we, here, attribute to Jeans as well) ought to be workable for 
a wide class of physical problems, classical as well as quantum.  But he was 
unable to rigourously justify every step, such as approximate ergodicity.
The breakthrough model of Ford--Kac--Mazur is what Khintchine was lacking.

But meanwhile, Professor von Plato, of Helsinki University, well versed in 
issues of logic, foundations, and axiomatics, was from another direction 
solving the problem of a logically unexceptionable, physically based 
definition of probability.  
It assumes the dynamics is ergodic and thus 
does not make use of the two-scale aspect of Statistical Mechanics, but it 
can be generalised to apply to approximately ergodic sequences as above.

More importantly, it can be generalised to quantum mechanics, where the 
dynamics is not even approximately ergodic.  The point is that if it is 
given that a system is in a certain subset of states, `normal' is a 
relative term with no probabilistic significance, we could instead 
simply study the statistical properties of \it that subset of states\rm.
If we knew that the system $M_{10^{23}}$ were in a definite state $v_o$, 
and if we could somehow justify that its statistical properties could still 
be approximately calculated by passing to the limit, where stochasticity 
arises, the essential logical structure of Statistical Mechanics does not 
change, only its technical calculation techniques may need adjustment.
Therefore, the thermodynamic limit of deterministic systems, all of them 
in a known pure state, could still be a stochastic process.  This is exactly 
what will happen in the case of a quantum amplifier.  So for our purposes, 
the full generality of approximate ergodicity and how to alter Professor 
von Plato's theory is not needed.  If a phase space consists of one or 
two points, ergodicity is not hard to investigate.  In the negative 
temperature situation, our limit system will consist of two points.
The interesting question of adapting Professor von Plato's theory to the logical 
structure of classical Statistical Mechanics, as outlined by Wiener 
and Khintchine, has been the topic for a separate paper.$^{28}$

\vskip-4pt
Nevertheless, we will recapitulate Professor von Plato's major breakthrough
in just enough detail so that our adaptation of it in the penultimate section 
of this paper, to quantum amplifiers, will be seen in context as a modest 
step further instead of a strange and unmotivated hack.
\vskip-4pt
\centerline{\bf Professor von Plato's Ergodic Theory of Probability}

\vskip-4pt
The Boltzmann-Gibbs development of statistical mechanics, especially as 
reformulated by Einstein in his doctoral thesis and subsequent work, 
does not quite follow the Jeans--Darwin--Fowler--Wiener--Khintchine 
logical structure as we have sketched it.  Primarily because it assumes 
ergodicity, and secondarily because it imposes a probability distribution 
on the phase space by \it fiat\rm\ instead of defining probability 
explicitly.  Nevertheless, in a well-known paper$^{20}$
Professor von Plato showed, in the Boltzmann--Gibbs--Einstein framework, 
how to give a logically unexceptionable, non-circular, and physically 
based definition of probability if the underlying dynamics is deterministic 
and ergodic.  In this section we recapitulate his logical structure.

Fix $n$ and suppose that the dynamical system $M_n$ is ergodic.  Suppose 
that $f_n$ is a fixed dynamical variable on $M_n$.  By Birkhoff's strong 
ergodic theorem, for almost all initial conditions  $x_o \in M_n$,  
the time average of $f_n$, $$\langle f_n(x_o)\rangle _t = \lim
_{T\rightarrow\infty} {1\over 2T} \int_{-T}^T f_n(x_t) dt $$
exists.  By the way, the theorem shows it almost always exists even if 
the dynamics is not ergodic, as long as the measure $d\mu(x)$ is preserved by the 
flow $x \mapsto x_t$.

Furthermore, it almost always equals the phase average of $f_n$, 
$$\langle f_n \rangle = \int_{M_n} f_n(x) d\mu(x).$$

Because mathematical probabilities of any event can be reconstructed if all 
the mathematical expectations of dynamical variables are known (provided, of course, that 
they satisfy the necessary additivity properties), 
we will have defined physical probability if we have defined physical correlates 
to all these expectations.  That is, we are reduced to the problem of giving a 
logically unexceptionable definitiion of the physical expectation of $f_n$, since 
$f_n$ was arbitrary.  Every measurable subset $S$ of the dynamical system is 
physical, so we need to give a physical meaning to the expectation of $f_n$ on 
$S$.  It suffices to assume $f_n$ is the characteristic function of $S$ and give 
a logical definition to the expectation of $f_n$ on all of $M_n$.  Professor von 
Plato has provided this in a way that mimics Einstein's use of the concept of 
probability (expectation) in his work, and in a way that is intuitively close 
to the idea of frequency.  

We define the expectation of $f_n$ to be the time average $\langle f_n \rangle _t.$
If the dynamics were discrete and fairly repetitive, as if it were a softball pitching machine, this would 
\it be\rm\ the intuitive idea of limiting frequency.
But it \it is\rm\ a statistical notion which does not involve the idea of probability.

More precisely, we define the expectation of $f_n$ to be the value of this time 
average for normal trajectories, and we know that this exists and is unique by 
ergodicity.

\vskip-3pt
The break-through improvement over naive frequency theories is that we have a 
physical basis: this definition is not logically circular because it assumes there 
is a deterministic, ergodic dynamics.  There is still a negligible set of 
`violators' in a physical sense: if we define `measurement' to mean the physical 
system is in a definite trajectory with initial condition $x_o$ 
and measurement of the value of $f_n$ means $\langle f_n(x_o) \rangle _t$, 
then it would follow that if this point 
were in the measure zero set such that the limit 
either were `wrong' or did not exist, then 
the actual measurement would not be equal to its expected value, or would be 
indeterminate.  Then the result of the `experiment' would not verify the `theory.'
This is a \it good\rm\ feature of Professor von Plato's theory.  The difficulty of 
negligible but real `violators' ought to be present in any physical theory of 
probability, his achievment is to make it logically and axiomatically harmless:
it is now a feature of praxis, with a theoretical correlative as part and parcel 
of a well-posed logical structure.

\vskip-9pt
I have argued elsewhere$^{25,26,28}$ that three modifications are desirable in Professor 
von Plato's theory.  Firstly, probability (or, equivalently, expectations) 
should be defined as phase averages and time averages should model measurements.
What for him is a definition ought to be re-interpreted 
as a theorem that probabilities predict the results of measurements but cannot 
otherwise be directly observed.  Secondly, following the Wiener--Khintchine 
programmatic, we should embed $M_n$ and $f_n$ in a thermodynamic limit sequence, 
and only assume the usual thermodynamic hypotheses which yield approximate 
ergodicity for \it some\rm\ $f_n$.  Thirdly, the underlying deterministic 
dynamics should be that of Quantum Mechanics in the particular physical setting 
of a quantum amplifier.  Probabilities should not be expected to arise 
except when microscopic aspects are being amplified to macroscopic dimensions.

\vskip-9pt
Because we are defining probability to be a limit phenomenon, our definition of 
probability is an `unpacking' style definition like Russell's theory of 
description, rather than a straightforward definition like Professor von Plato's.
A statement such as `the probability that the result of a measurement of $Q$ 
will yield the result $\lambda$ is $p$' is unpacked into a system of statements 
which are not actually about results or $Q$, just as the proposition `The current
King of France is bald' is not about a (non-existent) King of 
France.  Our analysis of measurement is that it is an idealised thermodynamic 
limit.  Therefore, our analysis of this proposition is that it asserts that 
a negative temperature amplifier can be coupled to the microscopic system 
being `measured' in such a way that when this combined system is modelled 
by a thermodynamic limit approximation, the approximation yields a measure 
on a classical space of results and the measure assigns a measure of $p$ 
to the labelled result.  The physical significance of such statements is 
that the approximation is an approximation to a time average of a directly 
physically meaningful function on the actual, unapproximated, quantum 
phase space of the amplifier.  This will become clearer by example.  See 
also the discussion in$^{28}$.
Probability is a linguistic feature of an approximation procedure.
(If infinite systems \it relevantly coupled\rm\ to things being measured 
really exist in Nature, perhaps some of this indirection will be unnecessary.)

The next section gives a toy model, which is basically that of H.S. Green and 
many others, and the logical status of the three modifications will be clear 
by example.  Because of the conjectured robustness of the limiting procedures 
of Statistical Mechanics, the future progress of mathematics ought to be able 
to show that the features of this model hold good in a certain amount of 
generality.  But we use a two-level model as is usual in the discussion of 
lasers and amplifiers in quantum optics.

\centerline{\bf The Concrete Model}

The picture is of $n$ particles in a line.  Each one passes 
on its state to its rightmost neighbour, and the first one receives its state from the one
on the right end (this is the cyclicity).  This takes place in time $ {1\over n}$.  
(A more realistic, but still simplified and explicitly solvable, model of an 
amplifying apparatus is based on the Curie-Weiss phase transition${}^{27}$.  
However, they are not concerned with deriving the properties of observables 
from the first three axioms, but feel free to use all six axioms indiscriminately.
Nor do they care to define macroscopic precisely, but content themselves with 
showing that the off-diagonal elements in the density matrix decay, in the 
limit.  This does not escape Bell's criticisms\plainfootnote*
{\eightrm\baselineskip=8pt 
`The result (15) is [called by Hepp[11]] the ``rigourous reduction of the wave packet.''
If the ``local observables'' $Q$ (as distinct in particular from the ``classical 
observables'') are thought of as those which can in principle actually be 
observed, then the vanishing of their matrix elements between two states 
means that coherent superpositions of $\psi_+$ and $\psi_-$ cannot be 
distinguished from incoherent mixtures thereof.  In quantum measurement 
theory such elimination of coherence is the philosopher's stone.  For with 
an incoherent mixture specialisation to one of its components can be 
regarded as a purely mental act, the innocent selection of a particular 
subensemble, from some total statistical ensemble, for particular further 
study.'   Bell, [3].  This is the criticism of statistical mechanical proposed 
solutions which is an independent, and prior, issue to any discussion of 
von Neumann's so-called projection postulate.
}
, which we are especially concerned 
to meet.  The underlying physics behind this model and their model is the same 
as that of H.\ S.\ Green${^9}$, \it op.\ cit\rm.)
 
\vskip-9pt
From now on, we distinguish between measurement apparatus and amplifying apparatus.  The  
amplifying apparatus we study will be an explicitly given quantum system with $n$ degrees  
of freedom, $M_n$, modelled by a Hilbert Space $\Cal H_n$ and with a Hamiltonian $H_n$. 
It approximates more and more to a measurement apparatus as $n\rightarrow\infty$.  The  
measurement apparatus is a thermodynamic limit of $M_n$, denoted $M_\infty$, and is a  
classical dynamical system.  Its states are the equilibrium states of the thermodynamic  
limit, and are not described by wave functions, its state space is a symplectic  
manifold, not a Hilbert space, has no linear structure, superposition of states is a  
nonsensical undefined concept for it.  Classical mixtures of its states are possible,  
as always in classical Statistical Mechanics.  One can take the viewpoint that  
measurement apparatuses and processes are unphysical idealisations of the only processes  
that are physical, the amplifying processes.  This is a valid logical interpretation of  
the measurement axioms (even, after some contortions, the reduction of the wave packet) 
and it does not involve any change in the operationalisation of the concept of  
measurement.  In fact, it grounds in concrete calculations what used to be operationalised  
anyway without justification: the fact that an amplifying apparatus must be large before  
the measurement axioms are verified.  A one-atom device does not perform a measurement\dots  
or reduce the wave packet. 
But if infinite systems \it do\rm\ exist in Nature, this concrete calculation should 
yield accurate results.  
Because it makes solving the problem harder, not easier, we will assume that 
only finite systems are realisable.  But the line of argument we introduce here is extensible and 
portable.  (If new systems, Hamiltonians, even forces of Nature are discovered 
that still fit into the logical framework of Quantum Mechanics, similar procedures 
to what we introduce here should carry over and give similar results.)  
 
Let the state space of an incident particle be $\bold C^2$.  This space has 
basis $\{\psi_0, \psi_1\}$.   For each $n$,  
$\Cal H_n$ is the Hilbert space of wave functions describing the state space of an  
$n$-oscillator system which is an amplifying device.   
We let $\Cal H_n=\oversetbrace {(n)} \to {\bold C^2\otimes\bold C^2\dots\bold C^2}$.   
 
 \vskip -.14in   In the presence of an incident particle in the state $\psi_1$, the amplifying apparatus will  
 evolve in time under the influence of $A_n$ (called ``Ming,''since it leads to a bright and clear phenomenon), a  cyclic 
nearest-neighbour interaction which is  
meant to model the idea of stimulated emission or a domino effect.  In the absence of a  
detectable particle, 
 the dynamics on the amplifying device will be trivial.   
This means that  
 
\vskip -.14in $\Cal H_n^{com}=(\bold C\cdot  \psi_1 \otimes \Cal H_n) \bigoplus (\bold C\cdot \psi_o \otimes \Cal H_n) 
 \text {\ \ and we put \ \ } H_n^{com}=I_2\otimes A_n + I_2\otimes I_{2^n}.$  
The intuition is that $\psi_1$ means the particle is in the state which the apparatus is designed to detect.   
 but $\psi_o$ means the particle is in a state which the apparatus is designed to ignore.

\vskip -.16in It will simplify things if we assume $n$ is prime.  (The general case can be reduced to  
this by perturbation.) 
Let $q={2^n-2\over n}$ (which is an integer by Fermat's little theorem). 
 
\vskip -.07in Let $i$ be any integer between 0 and $2^n-1$.  There are $n$ binary digits $d_i$ with  
$i=\sum_0^{n-1} 2^{i}d_i$ and uniquely so.  If $\{\vert 1\rangle,\vert 0\rangle\}$   
is a basis for $\Cal H_1$, then $\vert i\rangle_n= 
\otimes_0^{n-1} \vert d_i\rangle $ form a basis of $\Cal H_n$.  The intuition is that the  
$i^{th}$ oscillator is in an excited state $\vert 1 \rangle$ if $d_i=1$ and is in the  
ground state $\vert 0\rangle$ if $d_i=0$.  We also write  
$\vert i \rangle_n=\vert d_0 d_1 \dots d_{n-1} \rangle$. 
 
We wish to find Ming such that in one unit of time the  
$d_i$ are cycled as follows: 
 
\vskip -.46in $$e^{\frac{2\pi} h  A_n} \vert i \rangle_n = \vert d_{n-1} d_0 d_1 \dots d_{n-2} \rangle.$$

\vskip -.28in Choose a set of representatives $b_i$ such that every integer $k$ from 1 to $2^n-1$ can be written uniquely as 
$b_i2^m \mod (2^n-1)\bold Z$ for some $1\leq i \leq q $ and $0\leq m\leq n-1$, that is, $k=b_i2^m +j(2^n-1)$ for some $j$ 
but $i$ is unique.  (Since $2^n-1 $ and $2^m$ are relatively prime, no matter how $k$ and $m$ are fixed, there exist  
unique $b_i$ and $j$ satisfying this.)   
 
Then each $|b_i\rangle$ represents an orbit under the action of $e^{-2\pi  A_n}$.  Re-order the basis as follows: 
let $v_o=|b_1 \rangle$, $v_2=|b_12\rangle$, $v_2=|b_12^2\rangle$, \dots $v_{n-1}=|b_12^{n-1}\rangle$, 
$v_n=|b_2\rangle $,  
$v_{n+1}=|b_22\rangle$, \dots, $v_{2n-1}=|b_22^{n-1}\rangle$, $v_{2n}=|b_3\rangle$, etc., up to 
$v_{(q-1)n}=|b_q\rangle$, 
$v_{(q-1)n+1}=|b_q2\rangle \dots $,  
$v_{(q-1)n+n-1}=|b_q2^{n-1}\rangle$,  
but $(q-1)n+n-1=2^n-3$, so we have $2^n-2$ basis vectors accounted for.  Let $v_{2^n-2}=|0\rangle$ and 
$v_{2^n-1}=|2^n-1\rangle.$ 
 
Let $V_1$ be the space spanned by $\{v_o,\dots,v_{n-1}\}$, let $V_2$ be the space spanned by $\{v_n,\dots,v_{2n-1}\}$, 
etc., up to $V_q$.  Let $V_o$ be the space  
spanned by $\{v_{2^n-2}, v_{2^n-1}\}$.  
The Ming Hamiltonian operator $A_n$ on $\Cal H_n$ is a direct sum of its restrictions to the $V_i$.  Its restriction to $V_0$ 
is to be the zero operator.  Each $V_i$ is isomorphic to $V_1$ and we give the matrix of each restriction of $A_n$ with 
respect to the given bases. 
$${\text {\hskip -1in Solving \quad \ \ \quad\quad\quad   \quad \quad \quad }}  
{A_n}=\frac h{2\pi}\log \pmatrix  
0 & {0}& \dots&{0}& 1\\ 
1&{0} & \dots&{0}&{0} 
\\ 0&1&\ddots&{\vdots}&\vdots 
\\ \vdots&\ddots&{\ddots}&0&\vdots 
\\  0 &{\dots}&0 & 1&0\endpmatrix  , $$  
we obtain a cyclic skew-hermitian matrix, 
whose $i,j^{th}$ entry,  
${-ih\over n^2}\sum_{k=0}^{n-1} ke^{{2\pi i \over n}k(i-j)}$, 
 is approximately (if $n$ is large compared to $i-j$) 
$ih{(i-j)^{-1}\over2\pi} {\text {\ \ unless\ \ }} i=j {\text{\ \ in which case \ }} {ih\over2}$.   
 
     As usual in classical statistical mechanics, the observables are all abelian, and are given by measurable functions on the 
phase space, hence $f_n$ is an observable if  
$f_n:\Cal H_n^{com} \rightarrow \bold R$ is measurable and $f(c\psi)=f(\psi)$ for $c\in \bold C^{\times}$.   
In order to avoid confusion with the orthodox primitive concept of observable, modelled 
by a linear operator, we will not refer to $f_n$ as an observable and will not use the 
term `observable' in our system at all.  These measurable functions are dynamical 
variables, as usual in Hamiltonian dynamics.
 
\vskip-4pt
The intuitive picture is that this device is getting more and more classical as $n$ goes  
to infinity.  So the energy levels get closer and closer, approaching a continuum, the  
oscillators get closer and closer which is why the interaction, at a constant speed,  
travels from a oscillator to its neighbour in less and less time. 
If we adjusted  
by rescaling the dynamics to accomplish this, the entries of $H_n$ would diverge with $n$. 
 We rescale $h$ instead, so that it  
decreases as $\frac 1n$,
  This is typical of rescaling procedures in classical statistical mechanics.  It is physically meaningful because the 
thermodynamic limit is never physically real, it is  
only one of the $\Cal H_n$ which is physically real: $n$ is not a physical variable, it is a parameter.  Passing to the limit is only a 
mathematical convenience to obtain simple  
approximations for the physical truth about $\Cal H_{1.1\times 10^{36}}$.   
Since $h$ is truly small, this yields valid approximations. 

\vskip-4pt
We must couple the amplifier to the incident particle.  
The Hilbert space of the combined system is ${\Cal H}_n^{{\text com}}={\bold C}^2\otimes{\Cal H}_n
=<\psi_0> \otimes {\Cal H}_n \oplus <\psi_1> \otimes 
{\Cal H}_n$.  So we need only define $H_n^{{\text com}}$, the Hamiltonian of the joint system, by 
giving it on the first factor, where it is trivial, and on the second factor, where it is
$H_n$.  This is the explicit toy model of a quantum amplifier, we have now to study the 
question, what quantities at each finite stage correspond, in the limit, to a macroscopic 
pointer position of a classical measurement apparatus?

We need a precise notion of what is a macroscopic system, in terms of the axioms of 
quantum mechanics, and what is a macroscopic observable. 
Although our model owes a great deal to the Coleman--Hepp model, this notion is the exact 
opposite of Hepp's notion of a local observable (which is also the usual one in the infinite
volume thermodynamic limit of Haag, Ruelle, and others$^{21}$).  We wish to implement the intuition 
of a function on the phase space which cannot distinguish between states which differ from 
each other in a finite or negligible number of spots.  

In the classical methods of thermodynamics, one worked essentially with one 
observable at a time and there was really a sequence of them for 
each $n$.  I.e., for each $n$, one has $f_n$ a physically significant phase function on 
the space ${\Cal H}_n^{{\text com}}$, or its classical analogue.  And the physical significance is the 
same as $n$ varies.  It could be total energy of a part of the system, for example the 
Brownian mote.  For us, it could be a formalisation of some intuitive idea such as, the 
percentage of excited particles in the left half of the device.  
These methods are acutely, if disparagingly, described by R. Minlos$^{21}$.  
``For a long time the thermodynamic limit was understood and used too formally: the mean values of some 
local variables and some relations between them used to be calculated in a finite ensemble and then, in the formulas obtained, the limit passage was carried out.''
Excellent agreement with experimental results were obtained that way.
At any rate, we formally define such a sequence of $f_n$ to be macroscopic if whenever the 
sequence of norm one vectors $v_i\in{\bold C}^2$, $i>0$, satisfies 
\vskip-24pt
$$\lim_{n\rightarrow\infty} f_{n+n_o}(v_o\otimes v_{n_o+1}\otimes v_{n_o+2}\otimes\dots\otimes v_{n_o+n})$$
\vskip-9pt \noindent exists for some $n_o$ and some $v_o\in{\Cal H}^{{\text com}}_{n_o}$, then it is independent of the choice of
$n_o$ and $v_o$.  

We now define the family $f_n$, which in the limit, becomes the pointer position of the 
measuring apparatus.  There is a basis of ${\Cal H}_n^{{\text com}}$ consisting of separable vectors
 of the form $\psi_{\pi_0}\otimes\vert \pi_1 \pi_2 \dots\pi_n\rangle $ where, as before, the 
$\pi_i$ are 0 or 1.  Let $C$ be the set of basis vectors such that all but a negligible 
number of the $\pi_i$ for $i<n/2$ are 1 and all but a negligible number of the others are 0.
(This is the device being `cocked' and ready to detect.  It is very far from being a 
stable state, in the limit.)  By negligible, we mean that as a proportion of $n$, it 
goes to zero as $n$ increases.  For $v_n$ any state of the combined system, and we may 
take $v_n$ normalised of length unity,  let $c_i$ be the Fourier coefficients of $v_n$ with 
respect to the cocked basis vectors, i.e., those in $C$.  Define $f_n(v_n)=1-\sum_i|c_i|^2$.

\centerline{\bf Why $f_n$?  }

\vskip-10pt
We now discuss the physical basis for this choice.  Why a sum of amplitudes, 
squared, of the Fourier coefficients? (Up to an additive constant.)  
Opinions are divided as to the physical reality of the wave function.  
Certainly it is not easy to directly measure the wave function or quantum 
state of a system, even such a simple system as a harmonic oscillator.  
What are reasonable to measure are the amplitudes $\vert c_1\vert ^2$ of 
the state.  In fact these are 
the second main object of laboratory measurements (the first is the 
eigenvalues of $H$, of course).  It is not directly relevant that the 
interpretation of what one is doing when one experimentally determines 
these amplitudes is thus and such.  It is done every day with great 
experimental success.  The procedure is a little indirect, but not 
unduly so.  If we can reliably prepare isomorphic systems repeatedly 
in the same state,---now the current experiments on teleportation 
do this as a matter of course---, then repeated measurements of 
trials or measurements of observables give us the relative frequencies 
we need to estimate these amplitudes.  

\vskip-9pt
It has been often argued by many that this is too indirect, especially because 
any finite process would only achieve the determination of a finite 
number of these amplitudes.  And that therefore the wave function is 
not real, it is a probability wave, not a real physical state or a 
real physical object.  But this line of argument is philosophical, not 
physical.  The philosophy of positivism does indeed maintain the thesis 
that the meaning of a concept is \it how\rm\ one would observe it.  And 
Heisenberg and Mach did indeed maintain that theoretical physics should 
not use theoretical concepts that were meaningless, in this precise 
sense.  But Feynman, to take only one example,\plainfootnote{*}{\baselineskip=8pt \eightrm
Another is Weinberg, \eightit op.\ cit.\eightrm, pp.\ 178ff.} has argued against this 
extreme form of positivism.\baselineskip=8pt\plainfootnote\dag{\baselineskip=8pt \eightrm
``Another thing that people have emphasized since quantum mechanics was 
developed is the idea that we should not speak about those things that 
we cannot measure.\dots Unless a thing can be defined by measurement, 
it has no place in a theory.\dots The idea that this is what was the 
matter with classical theory \it is a false position\eightrm. \dots Just 
because we cannot \eightit measure\eightrm\  position and momentum precisely does 
not \eightit a priori\eightrm\ mean that we \eightit cannot\eightrm\ talk about them.  It 
only means that we \eightit need\eightrm\ not talk about them.
\dots
It is always good to know which ideas cannot be checked directly, but
it is not necessary to remove them all.  It is not true that we can pursue
science completely by using only those concepts which 
are directly subject to experiment.''
Feynman, Leighton, and Sands, \eightit Lectures on Physics\eightrm, vol. 3, Reading, Mass., 1965, pp. 2-8f.}   \baselineskip=24pt
Only disagreement with experiment, or 
logical incoherence, should tell against a physical theory, not 
philosophical strictures such as the positivist's.
(Even these are not always decisive: sometimes the logical incoherence 
has been fixed by theoreticians, and sometimes the disagreement with 
experiment has been fixed by better designed experiments\dots)

Since the squares can be measured, so can their sums, so this is a 
very natural function on the quantum phase space to study.
It is the very first abelian dynamical observable which would occur 
to anybody, and in fact it is the trace of a hermitian projection
operator, so in a different guise it has been much studied.

\centerline {\bf Passing to the Thermodynamic Limit: the Ming Effect}

Our procedure is modelled on that of Ford--Kac--Mazur, which indeed is modelled on the 
usual understanding of the Gibbs program.  However, a major physical difference is the 
renormalisation which we introduce.  Another is that the physical quantity whose 
limit they study is an autocorrelation; ours is a macroscopic pointer position.
Now we are interested in phenomena in the limit as $n$ approaches $\infty$, yet one cannot 
directly compare a value of $f_n$ on a vector $v_n$ with $f_{n+1}(v_n)$.  
In keeping with the procedures of classical statistical mechanics, 
one compares time or phase averages of the various $f_n$ as $n$ varies.  
(Phase averages would be taken over the submanifold of accessible states, it is more convenient
for us to deal with time averages.  Time averages have been made the basis for Professor von Plato's theory 
of the meaning of probability statements. 
Pauli$^{22}$ attributes the same idea to Einstein)
Let the incident particle
be in the state described by any (normalised) wave function in ${\bold C}^2$.  Let it be 
$v_0=a_0\psi_0+a_1\psi_1$.  The amplifier is in the state $|111\dots 000\rangle  $ in $C$.  
We now calculate the 
limit, as $n$ approaches $\infty$, of $<f_n>$, 
where $<f_n>$ means the time average of $f_n$ taken over a typical trajectory in the manifold 
of accessible states inside of ${\Cal H}_n^{{\text com}}$.

\vskip-4pt
We will then find a classical dynamical system $\Omega_\infty$ which has a mixed 
state $X$, which depends on $v_0$, and a classical dynamical variable $F$ whose 
expectation values match these limits.  At any rate, it is elementary to calculate 
$\lim_{n\rightarrow\infty} <f_n>$, it is $|a_1|^2$.
(The microscopic indident particle triggers a domino effect or macroscopic `flash' which is
`bright and clear' i.e., `Ming'.)

We search for $\Omega_\infty$, $F$, and $X_{v_0}$ as above, satisfying 
$$\int_{\Omega_\infty}FdX_{v_0}=\lim_{n\rightarrow\infty}<f_n>.$$
Let the state of the (classical limit) measurement apparatus where the pointer position 
points to cocked (and hence, an absence of detection) be the point $P_0$.  Let the state 
where the excited states of the apparatus are proceeding from out of its initial cocked 
state, and travelling steadily towards the right, be $P_1$.
Then $\Omega_{\infty}=\{P_0,P_1\}$.  The dynamical variables on this space are generated by
the characteristic functions of the two points, $\chi_{P_0}, \chi_{P_1}$. 
Let $F$ be $\chi_{P_1}$. It is the pointer position which registers detection.
The mixed state of  $\Omega_{\infty}=\{P_0,P_1\}$ which gives the right answer   
when the incident particle is in state $v_0$ is the probability distribution which gives 
$P_0$ the weight $|a_0|^2$, and $P_1$ the weight $|a_1|^2$.  
This is precisely what it means to say the the measuring apparatus will register the 
presence of the particle with probability $|a_1|^2$, and its absence with probability 
$|a_0|^2$.  

\centerline {\it Discussion of macroscopic}

\vskip-9pt
As Bell often remarked, there is no place in the physical world where one can 
put a precise cut between the microscopic and the macroscopic, and orthodox 
approaches to the foundations of Quantum Mechanics fail to be a logical 
axiomatisation for precisely this reason.  He never remarked that there 
was a difficulty in giving a precise definition of probability.  It has 
transpired that these two difficulties are connected.  We have made 
`macroscopic' a linguistic notion so that statements using it are seen to 
have reference to an approximation scheme rather than directly to elements 
of physical reality.  (The cut has been pushed out to infinity, \it i.e.\rm, 
there is no real cut).  The sticking point in all attempts to define 
probability which remain close to the frequency theory has always been what 
to say about the negligible set of violators.  In our re-deployment, 
the notion of macroscopic is that a phenomenon, such as pointer position, 
is macroscopic if it is defined after neglecting a negligible set of 
position violators.  For example, we defined a variable on the quantum phase 
space which was insensitive to `missing' parts of the needle pointer 
if their bulk was too small to be seen by the naked eye.  It's limit was, 
then, one that neglected it.  Because of the situation of amplification, 
there are two physical scales, so it is physically natural to introduce 
this kind of concept of negligibility.  So, what was a `sticking point' 
in other schemes, is now quite natural.  This answers Feynman's question 
as to why probabilities arise from the necessity of amplification.  If 
the naked eye did not neglect such things, there would be no need of 
amplification.  Since there is such a need, the naked eye does neglect 
them.  Neglecting `violators' seems to be the characteristic feature of 
frequency-base definitions of probability, and we have the same feature 
here too, only in a logically `clean' location.

\centerline{\bf Discussion of Paralipomena}

`Event' is not a concept that can be expressed within the fully formalised, 
adequately axiomatised language of Physics.  It has been left out of our short 
list of fundamental concepts and it is not defined.  (The list was: system, 
state, Hamiltonian, time.)\plainfootnote
* {\eightrm \baselineskip=8pt Compare this to Hertz's list, which was space, 
mass, time.}
Physically, there is something `noisy' about actual, finite size, 
event counters.  This suggests that `event' should not be made one of 
the fundamental primitive concepts, and it even suggests it should not
even be considered as exactly physical, but only an idealised 
approximation such as occurs in the thermodynamic limit.  `Event' 
seems to be a classical concept of limited validity, no wonder 
the probabilities arise when we start talking about `events'.  The word 
does not occur in the first three axioms of Quantum Mechanics, so 
we did not exert ourselves to introduce it.  It was a mistake for 
Born to try to interpret the wave function in terms of events.  
What we have done instead is to derive the Born rule as a useful 
approximation in certain physical situations.  This is not an interpretation.

  It is true that individual 
consciousnesses certainly experience `events' and, indeed, experience them 
singly and one at a time.  What we have succeeded in defining in terms of 
the fundamental axiomatic concepts of Physics is only some statistical 
properties of (suitably appropriate) sets of events.  Nature and Heisenberg 
have taught us that it is only these statistical properties which are 
replicable.  Now, a `result' is not an `experimental result' unless it is 
replicable.\baselineskip=8pt\plainfootnote*
{\baselineskip=8pt \eightrm Philosophical proof by laughter: every scientist immediately starts laughing 
when they hear the title of a satirical journal occasionally put out by 
scientists in America.  The title is, \eightit The Journal of Irreproducible Results.}
\baselineskip=24pt
  Therefore the requirement of agreement with experiment 
is satisfied by our theory, because all replicable statistical regularities 
of the experiences of events by individual consciousnesses can be stated 
precisely in the language of Physics, calculated, and the results do in 
fact agree (to within experimental accuracy which, \it inter alia\rm\ 
means neglecting negligible events as a matter of \it praxis\rm) with 
experiment.

In fact the following statement is also capable of precise formulation 
and is verified: each single event results in a definite result.
Note well that this statement is itself a statistical regularity! It is 
modelled by the feature of our formalism which says the thermodynamic limit 
system has precisely two points.  The only possible events are the points.
Not superpositions.  It is a classical system.

\vskip-6pt
No statements in the precise language of Physics are falsified by 
experimental results or by experience because, for example, the following 
colloquial statement cannot be translated into our system: 
(I am afraid this is very Bohrian but after all, Bohr was not actually 
very wrong) `I ran the experiment and found the spin was up.'
Within the axiomatised language of Physics, the concept of `experiment' is 
merely a statistical concept and only describes an ensemble.  No actual 
dynamical system and initial condition is `an experiment'.  It is an 
amplifier.  The concept of `measurement' has been defined as a limit 
concept.  Fortunately, the colloquial statement is also not replicable! 
Therefore, as Heisenberg taught us, it is not \it necessary\rm\ for 
a physical theory to account for it.

\vskip-6pt
But, `I ran the experiment and found a definite result' (or, what is 
the same thing, `I ran the experiment once and found that the spin was 
either up or down') is replicable.
It is modelled by saying that the correlation of the stochastic process 
on our limit (two-point) space \it with itself\rm \ is unity.

\vskip-6pt
There is a consensus that Quantum Mechanics is mysterious and un-understandable.
It might, therefore, be a mistake to propose a theory which eliminates all 
mysteries.  The point of the Hilbert problem is simply to shove the mystery 
into some \it other\rm\ location than the axiomatics and logical structure.
This paper has shown that the mystery can be transferred into the usual 
philosophical dualism between Physics (materialism) and subjectivity 
(Geist).  More particularly, there is no physical result which is actually 
violated by our experience.  Every contingency which can actually be stated 
clearly and precisely in our axiom system is `subject to the laws of 
Physics'.  This is a true dualism without being either a monotonous 
parallelism or a total disconnect.  But it is not a domain problem: every 
physically formulatable contingency is within the domain of the laws of 
Physics (except perhaps negligible contingencies?)  It is not as though 
we can experience something which `violates' the laws of Physics.

The disconnect is not total because in our experience we experience 
subjective regularities of certain sorts, and the language of Physics 
is capable of describing statistical regularities of the same sort, and 
the two languages do truly describe, in a concordant fashion, the 
same regularities.  So we have a dualism which is not a parallelism, 
but has a precise dictionary for some (but not all) concepts.  Perhaps 
other connections remain to be discovered.  This is the first such 
connection which is not a full parallelism to be precisely formulated, 
but by no means rules out more.  

We abandon `\it das sog.\ Prinzip vom psycho-physikalischen Parallelismus'\plainfootnote*
{\eightrm von Neumann, \eightit op.\ cit.\eightrm, p.\ 223.} 
\rm in favour of more modest 
functors between the two sides of the duality.

\centerline{\bf Summary}

     This answers a famous question of Einstein's partly
 affirmatively and partly negatively.  It is possible to 
derive the probabilities in measurement results from an 
underlying deterministic dynamics analogously to the way
 it was done in classical statistical mechanics of Einstein's day.  It is not necessary to assume that quantum
 mechanics is incomplete in order to do this: we may take
 the wave function as the complete description of nature 
and Schr\"odinger's equation as exactly and universally 
valid.

This analysis shows that wave-packet reduction in the strong topology, 
even approximately, need not occur.  
It also shows that the transmutation of quantum amplitudes into classical 
probabilities depends only on the macroscopic nature of the pointer position  
and its coupling to the microscopic system being measured.  It does not depend 
on any back-force being exerted on the microscopic system.  
The coupling can be as gentle, in its effect on the microscopic system, as desired.
It suggests that the degree of 
validity of `measurement,' as an approximation to a physical amplification 
process, depends on the size of the apparatus.  Mesoscopic 
amplifiers should, then, demonstrate detectable noise phenomena 
in comparison to macroscopic amplifiers.

     This analysis further shows that it is not necessary to invoke the effect 
of the environment in order to construct a logically coherent theory of 
decoherence.  The fact that probabilities arise even from an amplifier which is
 in a pure state shows that quantum measurement can be explained without 
super-selection rules.  Thus the question whether the observed behaviour of 
measurement processes is due to the coupling between the apparatus and the 
microscopic system, or due to the coupling between the apparatus and the 
environment$^{24}$, becomes a question for experiment.

\baselineskip=20pt
\centerline{\bf References}

\noindent [1] E. Wigner, Z. Phys.\ {\bf133} (1952), 101; Am.\ J. Phys.\ {\bf31} (1963), 6.

\noindent [2] J. Jauch, E. Wigner and M. Yanase, Nuovo Cimento {\bf48} (1967), 144.

\noindent [3] J. Bell, Helv.\ Phys.\ Acta {\bf48} (1975) 447; Physics World {\bf3} (1990), 33.

\noindent [4] C. Darwin and R. Fowler, Philos.\ Mag.\ {\bf44} (1922), 450; 823; Proc.\ Cambridge Philos.\ Soc.\ {\bf21} (1922), 391.

\noindent [5] A. Khintchine, {\it Mathematical Foundations of Statistical Mechanics}, Moscow, 1943.

\noindent [6] G. Ford, M. Kac and P. Mazur, J. Math.\ Phys.\ {\bf6} (1965), 504.

\noindent [7] J. Lewis and H. Maassen, {\it Lecture Notes in Mathematics} {\bf1055}, Berlin, 1984, 245.

\noindent [8] E. Farhi, J. Goldstone and S. Gutmann, Ann.\ Phys.\ (NY) {\bf192} (1989), 368.

\noindent [9] H. Green, Nuovo Cimento {\bf9} (1958), 880.  

\noindent [10] A. Daneri, A. Loinger and G. Prosperi, Nucl.\ Phys.\ {\bf33} (1962), 297.   

\noindent [11] K. Hepp, Helv.\ Phys.\  Acta {\bf45} (1972), 237.

\noindent [12] J. Schwinger, J. Math.\ Phys.\  {\bf2} (1961), 407.

\noindent [13] W. Zurek, Phys.\  Rev.\  D {\bf24} (1981), 1516; {\bf26} (1982), 1862; Physica Scripta {\bf76} (1998), 186.

\noindent [14] C. Gardiner and P. Zoller, {\it Quantum Noise}, Berlin, 2000, pp.\ 212-229.

\noindent [15] M. Collet, G. Milburn and D. Walls, Phys.\ Rev.\ D {\bf32} (1985), 3208. 

\noindent [16] J. von Neumann, {\it Mathematicsche Grundlagen der Quantenmechanik}, Berlin, 1932.

\noindent [17] P. Dirac, {\it Directions in Physics}, New York, 1978, p.\ 10.

\noindent [18] R. Feynman and A. Hibbs, {\it Quantum Mechanics and Path Integrals}, New York, 1965.

\noindent [19] A. Sudbery, {\it Quantum Mechanics and the Particles of Nature}, Cambridge, 1986, 41ff.

\noindent [20] J. von Plato, ``Ergodic Theory and the Foundations of Probability,'' in {\it Causation, Chance, and Credence, Proceedings of the Irvine Conference on Probability and Causation}, edited by B. Skyrms and W. Harper, vol.\ 1, Kluwer, 1988, pp.\ 257-277.

\noindent [21] R. Minlos, {\it Introduction to Mathematical Statistical Physics}, Providence, 2000, p.\ 22.

\noindent [22] W. Pauli, ``L'apport d'Einstein \`a la th\'eorie quantique,'' in, Schilpp, ed., \it Einstein\rm, Evanston, 1949.

\noindent [23] K. Hannabuss, Helv.\ Phys.\ Acta {\bf57} (1984), 610; Ann.\ Phys.\ (NY) {\bf239} (1995), 296.

\noindent [24] H. Zeh, in {\it Proceedings of the II International Wigner Symposium}, Goslar, Germany, 1991, edited by H. Doebner, W. Scherer, and F. Schroeck, Jr., World Scientific, 1993, 205.

\noindent [25] J. Johnson, Statistical Mechanics of Amplifying Apparatus, Paper presented at the VIII International Wigner Symposium, New York City, May 26-June 1, 2003.

\noindent [26] J. Johnson, in {\it Quantum Theory and Symmetries, Proceedings of the third International Symposium,} Cincinnati, 2003, edited by P. Argyres \it et al\rm., Singapore, 2004, 133.

\noindent [27] A. Allahverdyan, R. Balian, T. Nieuwenhuizen, 
Europhys.Lett. 61 (2003) 452-458

\noindent [28] J. Johnson, \it Probability as a Multi-Scale Phenomenon\rm, Paper presented at the Special Session on Mathematical Challenges in Physical and Engineering Sciences, at the 1017th meeting of the American Mathematical Society, 
Durham, New Hampshire, April 22-23, 2006.

\noindent [28] J. Johnson, \it The Logical Status of Probability Assertions\rm, submitted.
\end